\documentclass[twocolumn]{autart} 

\usepackage{lineno,hyperref}
\modulolinenumbers[5]

\usepackage[longnamesfirst]{natbib}

\usepackage{amsmath}
\usepackage{cite}
\usepackage{bbold}
\usepackage{amssymb}
\usepackage{graphicx}
\usepackage{enumerate}
\usepackage{comment,kotex,color}
\usepackage{setspace}

\newtheorem{thm1}{\bf Theorem}

\newtheorem{lem1}{\bf Lemma}

\newtheorem{rem1}{\bf Remark}

\newtheorem{property}{\bf Property}
\newtheorem{propertyp}{\bf Property}

\newtheorem{assumE}{\bf Assumption}
\newtheorem{assumW}{\bf Assumption}
\newtheorem{assumG}{\bf Assumption}

\newtheorem{assumM}{\bf Assumption}

\newcommand*{\QEDB}{\hfill\ensuremath{\square}}
\newcommand*{\QEDBB}{\hfill\ensuremath{\blacksquare}}

\begin{document}

\begin{frontmatter}

\title{Rapid and robust synchronization via weak synaptic coupling\thanksref{mytitlenote}}
\thanks[mytitlenote]{This paper was not presented at any IFAC meeting. 
Corresponding author R. Sepulchre.
The research leading to these results has received funding from the European Research Council under the Advanced ERC Grant Agreement Switchlet n.670645 and SpikyControl n.101054323.
This work was done while Jin Gyu Lee was with University of Cambridge.}
\vspace{-4mm}
\subtitle{\large Extended arXiv version}

\author[Inria]{Jin Gyu Lee}\ead{jin-gyu.lee@inria.fr},
\author[Cambridge]{Rodolphe Sepulchre}\ead{r.sepulchre@eng.cam.ac.uk}

\address[Inria]{Inria, University of Lille, CNRS, UMR 9189 - CRIStAL, F-59000 Lille, France}
\address[Cambridge]{Control Group, Department of Engineering, University of Cambridge, United Kingdom}

\begin{keyword}
Rapid synchronization, heterogeneous multi-agents, excitable systems, weak synaptic coupling, slow-fast models
\end{keyword}

\begin{abstract}
This paper examines how {\it weak} synaptic coupling can achieve {\it rapid} synchronization in {\it heterogeneous} networks. 
The assumptions aim at capturing the key mathematical properties that make this possible for biophysical networks. 
In particular, the combination of nodal {\it excitability} and {\it synaptic} coupling are shown to be essential to the phenomenon.
\end{abstract}

\end{frontmatter}

\section{Introduction}

Biophysical neuronal networks exhibit {\it rapid} transitions between asynchronous and synchronous modes in spite of {\it weak} coupling and significant {\it heterogeneity} in the network. 
Such properties might be desirable in engineered networks, but they do not seem immediately aligned with the mathematical analysis of synchronization. 
A standard model of synchronization in physical models is through diffusive coupling in a network of Van der Pol nonlinear oscillators, see, e.g.,~\citep{wang_slotine04,stan_sepulchre07,angeli02,arcak07}. 
Synchronization in those models is the result of a contraction property for the error dynamics. 
The analysis requires {\it identical} oscillators and the coupling strength must be {\it strong} enough to overcome the shortage of contraction in the decoupled model. 
The robustness of this mechanism to the heterogeneity of the network has been studied by several authors, see, e.g.,~\citep{kim2016robustness,panteley2017synchronization,Lee2018,jglee18automatica}. 
However, the common conclusion of such analysis is that stronger coupling is required to compensate for stronger heterogeneity. 
This conclusion is consistent with synchronization studies in more abstract phase models such as the celebrated Kuramoto model~\citep{kuramoto,strogatz,dorfler}. 

The goal of the present paper is to highlight the additional modeling properties needed to reconcile the apparent discrepancy between biophysical observations and mathematical analysis. 
We will analyze synchronization in a model of {\it excitable} neurons under {\it synaptic} coupling. 
Those two properties are central to biophysical models of synchronization: a mechanism akin to fast threshold modulation~\citep{somers1993rapid}. 
The nonlinear nature of the synaptic coupling makes it possible to bound the coupling by a given constant and at the same time make its gain high in a narrow amplitude range. 
Synaptic coupling hence reconciles the {\it weak} coupling biophysical assumption with the localized {\it strong} coupling property required for mathematical contraction. 
Excitable models have a {\it threshold}, that is, a highly localized amplitude range where the spike can be triggered.
Thanks to the threshold, the range of strong coupling can be narrowed down to where it is most beneficial to synchronization.
The two properties combined together drastically simplify the analysis of the overall network because the neurons are nearly decoupled almost everywhere in the state-space.
As a result, the convergence analysis decomposes into a sequence of distinct phenomena:
\begin{enumerate}
\item Rapid convergence of each individual neuron to a neighborhood of its limit cycle attractor, determined by the absolute bound on the coupling term treated as an exogenous disturbance; the weaker the coupling, the better the bound (Sections~\ref{subsec:spike}--\ref{subsec:syn}).
\item Owing to the fast-slow nature of the oscillations and to the synaptic nature of the coupling, the limit cycle trajectory of each neuron is rapidly entrained by the short coupling pulses triggered by each spike of presynaptic neurons. 
The rate of entrainment is independent of the strength of the coupling (Section~\ref{subsec:local}).
\item {\it Phase} synchronization of the spiking events occurs because the coupling pulses of the presynaptic neurons only affect the phase of the postsynaptic neuron near the threshold, the narrow range where they reduce the phase difference (Section~\ref{subsec:global_convergence}).
\end{enumerate}

Our objective is to show that those three intuitive properties can be demonstrated in a standard mathematical model using standard analysis. 
In particular, the simple linearization analysis of the differential equation highlights how the synaptic coupling provides strong diffusive coupling only near the threshold, making the overall network nearly decoupled anywhere else. 

The specificity of {\it pulse} coupling with respect to {\it diffusive} coupling has been thoroughly studied in the dynamical systems literature, see, e.g., the seminal paper~\citep{mirollostrogatz91} and the tutorial paper~\citep{mauroy2012kick}. 
However, combining the slow-fast nature of spiking oscillations with a reduction to phase models has proven challenging~\citep{izhikevich2000phase,sacre2014sensitivity,sacre2015singularly}. 
Those technical obstacles have made it difficult to connect the literature on diffusive coupling in ordinary differential equations with the literature on abstract phase models. 
The analysis in the present paper does not perform any phase reduction and proceeds straight from linearization analysis of the full differential equation model: we instead perform a network-wise singular perturbation analysis. 

\begin{figure}[h]
\begin{center}
\includegraphics[width=\columnwidth]{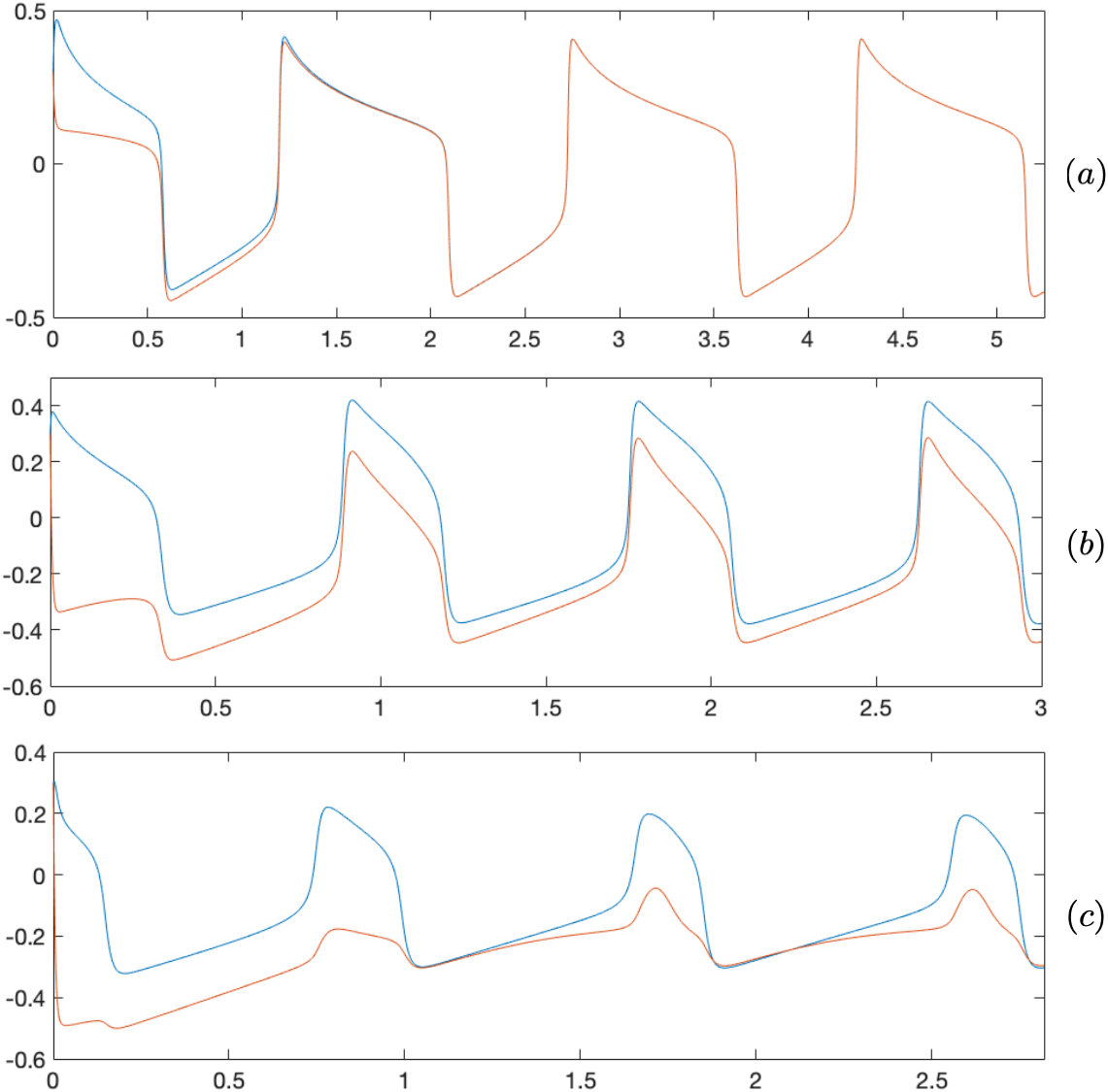}
\caption{Rapid synchronization of two identical (a) and non-identical (b) excitable systems under weak excitatory synaptic coupling.
(c) Poor synchronization of the same non-identical excitable systems under strong diffusive coupling.}
\label{fig:1}
\end{center}
\end{figure}

A strong source of inspiration for the present paper is the work by Somers and Kopell~\citep{somers1993rapid,somers1995waves}. 
To the best of our knowledge, they were the first to observe the rapid synchronization phenomenon in a model that combines excitability and synaptic coupling (which yields fast threshold modulation).  
They provided a detailed analysis of two identical relaxation oscillators (two-time scale oscillators characterized by a square-wave oscillation) under weak excitatory synaptic coupling and provided numerical evidence that a ring of relaxation oscillators synchronizes quicker than a ring of sinusoidal oscillators, under excitatory synaptic coupling.  
This basic but key observation is illustrated in Figure~\ref{fig:1} (a) and is a starting point for the analysis in the present paper.

The phase portrait of Figure~\ref{fig:1} (a) illustrates why localizing the coupling near the threshold is key to rapid synchronization.
In this example, the vertical direction is the direction where coupling is exerted.
Under uniform diffusive coupling, most of the coupling force is balanced by the strong attraction of individuals to their own limit cycles, hence resulting in a slower convergence rate for synchronization (Figure~\ref{fig:1} (c) illustrates that the strong diffusive coupling significantly deforms the trajectory of the oscillation).
In contrast, the synaptic coupling is active only near the threshold, where it matches the direction of pulling.
This mechanism is quite robust to heterogeneity as illustrated by Figure~\ref{fig:1} (b).
An idea to design nonlinear coupling to synchronize identical agents efficiently under the same philosophy has been introduced in~\citep{pavlov2022nonlinear}, where the goal is to make the error dynamics contractive everywhere.

The remainder of the paper is organized as follows.
We motivate the discussions made earlier with a concrete example in Section~\ref{sec:me}.
Then, we introduce a generalized model in Section~\ref{sec:model}.
The latter two steps of the convergence analysis are detailed in Section~\ref{sec:main} based on a network-wise reduction illustrated in Section~\ref{sec:i_n_f} followed by a discussion in Section~\ref{sec:main_bio}.
The first step of the convergence is straightforward from the definition of the model (Section~\ref{subsec:syn}).
A simulation result illustrating the robustness of these biophysical assumptions with respect to heterogeneity is given in Section~\ref{sec:sim}.
Finally, we conclude in Section~\ref{sec:conc}.
Additional technical details are given in the Appendix.

{\em Notation:}
A (directed) graph is a pair $\mathcal{G} = (\mathcal{N}, \mathcal{E})$ consisting of a finite nonempty set of nodes $\mathcal{N}$ and an edge set of ordered pairs of nodes $\mathcal{E} \subseteq \mathcal{N} \times \mathcal{N}$, where $(i,i)\notin\mathcal{E}$ for all $i\in\mathcal{N}$ (i.e., the graph does not contain self-loops).
The set $\mathcal{N}_{i} = \{j\in \mathcal{N}:(j,i) \in \mathcal{E}\}$ denotes the neighbors of the node~$i$.
A tuple $(i_{0}, i_{1}, ..., i_{l})$ is called a \emph{path} (of length $l$) from node $i_0$ to node $i_l$ if $i_{k} \in \mathcal{N}_{i_{k+1}}$ for all $k=0,\ldots,l-1$.
Given a graph $\mathcal{G} = (\mathcal{N}, \mathcal{E})$, let $\mathcal{N}' \subseteq \mathcal{N}$ and $\mathcal{E}'\subseteq \mathcal{E}|_{\mathcal{N}'} := \{(j,i)\in\mathcal{E}:i,j\in\mathcal{N}'\}$.
Then the pair $\mathcal{G}'=(\mathcal{N}',\mathcal{E}')$ is called a \emph{subgraph} of $\mathcal{G}$.
If $\mathcal{N}'=\mathcal{N}$, then $\mathcal{G}'$ is a spanning subgraph.
A graph $\mathcal{G}$ is \emph{strongly connected} if, for any pair of agents $i$ and $j$, there exists a path from $i$ to $j$.
A spanning subgraph $\mathcal{T}$ of $\mathcal{G}$ having a root $i$ such that there is exactly one path from $i$ to any other node is called a \emph{spanning tree} of $\mathcal{G}$.
An independent strongly connected component (iSCC) of $\mathcal{G}$ is a subgraph $\mathcal{G}' = (\mathcal{N}', \mathcal{E}')$ such that it is maximal subject to being strongly connected and satisfies that $(j, i)\notin \mathcal{E}$ for any $j \in \mathcal{N} \setminus\mathcal{N}'$ and $i \in \mathcal{N}'$.
For vectors or matrices $a$ and $b$, ${\rm col}(a,b) := [a^T,b^T]^T$.

\section{A motivating example}\label{sec:me}

The simulation in Figure~\ref{fig:1} involves the standard Morris-Lecar excitable model 
\begin{align}\label{eq:mot_ex}
\begin{split}
\tau(v)\dot{x} &= -x + x_\infty(v),  \\
\epsilon\dot{v} &= g_\text{L}(-0.4 - v) + g_\text{Ca} m_\infty(v)(1 - v) \\
&\quad + g_\text{K} x(-0.7 - v) + 0.4 + I
\end{split}
\end{align}
where the small parameter $\epsilon = 0.02$ controls the time scale separation of the model, and
\begin{align*}
\tau(v) = \frac{1}{\cosh\left(\frac{v+0.1}{0.29}\right)}, \,\,\, x_\infty(v) &= \frac{1}{2} + \frac{1}{2}\tanh\left(\frac{v+0.1}{0.145}\right), \\
 m_\infty(v) &= \frac{1}{2} + \frac{1}{2}\tanh\left(\frac{v}{0.15}\right).
\end{align*}

Figure~\ref{fig:1} (a) considers two identical neurons, with (voltage) states $v$ and $\hat v$. 
The numerical values of the leak, $\text{Ca}^{++}$, and $\text{K}^+$ maximal conductances are $g_\text{L}=0.5$, $g_\text{Ca}=1$, and $g_\text{K}=2$, respectively. 
The coupling term from the (presynaptic) neuron $\hat v$ to the (postsynaptic) neuron $v$ is the synaptic current 
$$ I = g_\text{coup} (1 - v) m_\infty(\hat{v})  $$
called {\it excitatory} for $v \le 1$ and {\it weak} if the maximal conductance parameter $g_\text{coup}>0 $ is small. 
The simulation uses $g_\text{coup} = 0.67$. 
We note that the weaker $g_\text{coup}$, the weaker the external input $I$ in~\eqref{eq:mot_ex}.  
We assume that the coupling is symmetric, just for simplicity.

A simple calculation provides the linearized error dynamics along the synchronized solution $v(\cdot) = \hat{v}(\cdot)$. 
The linearized error dynamics have the following expression:
\begin{align}\label{eq:mot_linear}
\begin{split}
\dot{\delta\tilde{x}} &= -\frac{1}{\tau(v(t))}\delta\tilde{x} + b(t) \delta\tilde{v}, \\
\epsilon\dot{\delta\tilde{v}} &= \left[a(t) - k(t)\right]\delta\tilde{v} + g_\text{K}(-0.7-v(t))\delta\tilde{x},
\end{split}
\end{align}
with 
\begin{align*}
b(t)&= \frac{\tau'(v(t))\left[x(t) - x_\infty(v(t))\right] + \tau(v(t))x_\infty'(v(t))}{\tau(v(t))^2}, \\
a(t) &= -g_\text{L} - (g_\text{Ca} + g_\text{coup})m_\infty(v(t)) \\
&\quad + g_\text{Ca} m_\infty'(v(t))(1 - v(t)) - g_\text{K}x(t), \\
k(t) &= g_\text{coup} (1 - v(t)) m_\infty'(v(t)).
\end{align*}
System~\eqref{eq:mot_linear} has the familiar form of error dynamics of a second-order linear system with diffusive coupling.
The coupling gain $k(t)$ is always non-negative and can be large in the local range of the sigmoidal nonlinearity, even if the synaptic coupling strength $g_\text{coup}$ is small.
In other words, the synaptic coupling can provide a {\it strong} diffusive coupling gain in a {\it localized} voltage range.
According to the classical theory of diffusive coupling, this large gain can make the linearized error dynamics contractive, which guarantees the exponential synchronization of solutions starting in the vicinity of the limit cycle.
In the rest of the paper, we will see that this results in an open set of synchronous spiking in the limiting case in which the spikes take place infinitely fast: that is akin to fast threshold modulation, yielding rapid convergence.

It will be also apparent that owing to the spiky nature of the oscillation, the coupling will almost vanish along the slow branch of the limit cycle attractor, only contributing to the contraction during a synchronous spiking with the help of the sufficiently steep synaptic coupling and the sufficient time scale separation, which gives nodal excitability.

This will be not so different for heterogeneous systems as illustrated in Figure~\ref{fig:1} (b). 
Here, we consider the heterogeneous network where the second neuron has the maximal conductance parameters $g_\text{L}=0.25$, $g_\text{Ca}=0.5$, and $g_\text{K}=4$. 
According to the blended dynamics theory~\citep{kim2016robustness,panteley2017synchronization,jglee18automatica}, a sufficiently large gain $k(t)$ can also make the error dynamics of {\it heterogeneous} networks contractive. 

For the numerical example shown in this section, the value
\begin{align*}
a(t) - k(t) &= - 0.5 - (1 + g_\text{coup})m_\infty(v(t)) \\
&\quad + (1 - g_\text{coup})m_\infty'(v(t))(1 - v(t)) - 2x(t) 
\end{align*}
becomes uniformly negative when (since $x \in [0, 1]$)
\begin{align*}
g_\text{coup} &= 0.67 \\
&>  \max_{v \in [-0.7, 1]} \frac{-0.5 - m_\infty(v) + m_\infty'(v)(1-v)}{m_\infty(v) + m_\infty'(v)(1-v)}.
\end{align*}
Meanwhile, further simulation with a smaller $g_\text{coup}$ suggests an example where the periodic linear system is contracting over one period, but not point-wise in time. 
In the general model treated in this paper, this mechanism will lead to synchronization of the {\it spiking times}, rather than arbitrary precision approximate synchronization to a common limit cycle trajectory, especially for heterogeneous networks.
The spiking time refers to the timing of the fast time scale behavior.
This timing shrinks to a single instant of time in the singular limit where the spikes occur infinitely fast.
See Section~\ref{sec:model} for more detail.
This weaker form of synchronization (called {\it phase} synchronization) is the one considered in this paper. 

\section{Model assumptions}\label{sec:model}

\subsection{Spiking model for individual neurons}\label{subsec:spike}

We consider $N$ heterogeneous neurons individually modeled by the second-order ultrafast-fast-slow system
\begin{align}\label{eq:ind_neu}
\begin{split}
\tau_{i, \epsilon}(v_i)\dot{x}_i &= -x_i + x_{i, \infty}(v_i) \\
\epsilon\dot{v}_i &= f_i(x_i, v_i) + (\overline{E}_i - v_i)G_i
\end{split}
\end{align}
with $i \in \mathcal{N} := \{1, \dots, N\}$, and $G_i$ is the external input of neuron $i$.
Here, $\tau_{i, \epsilon}(v) > 0$ is a state-dependent fast-slow time constant.
In the absence of coupling, the external input is $G_i = 0$. 
The phase portrait is then the classical Morris-Lecar phase portrait.

\begin{figure}[h]
\begin{center}
\includegraphics[width=0.99\columnwidth]{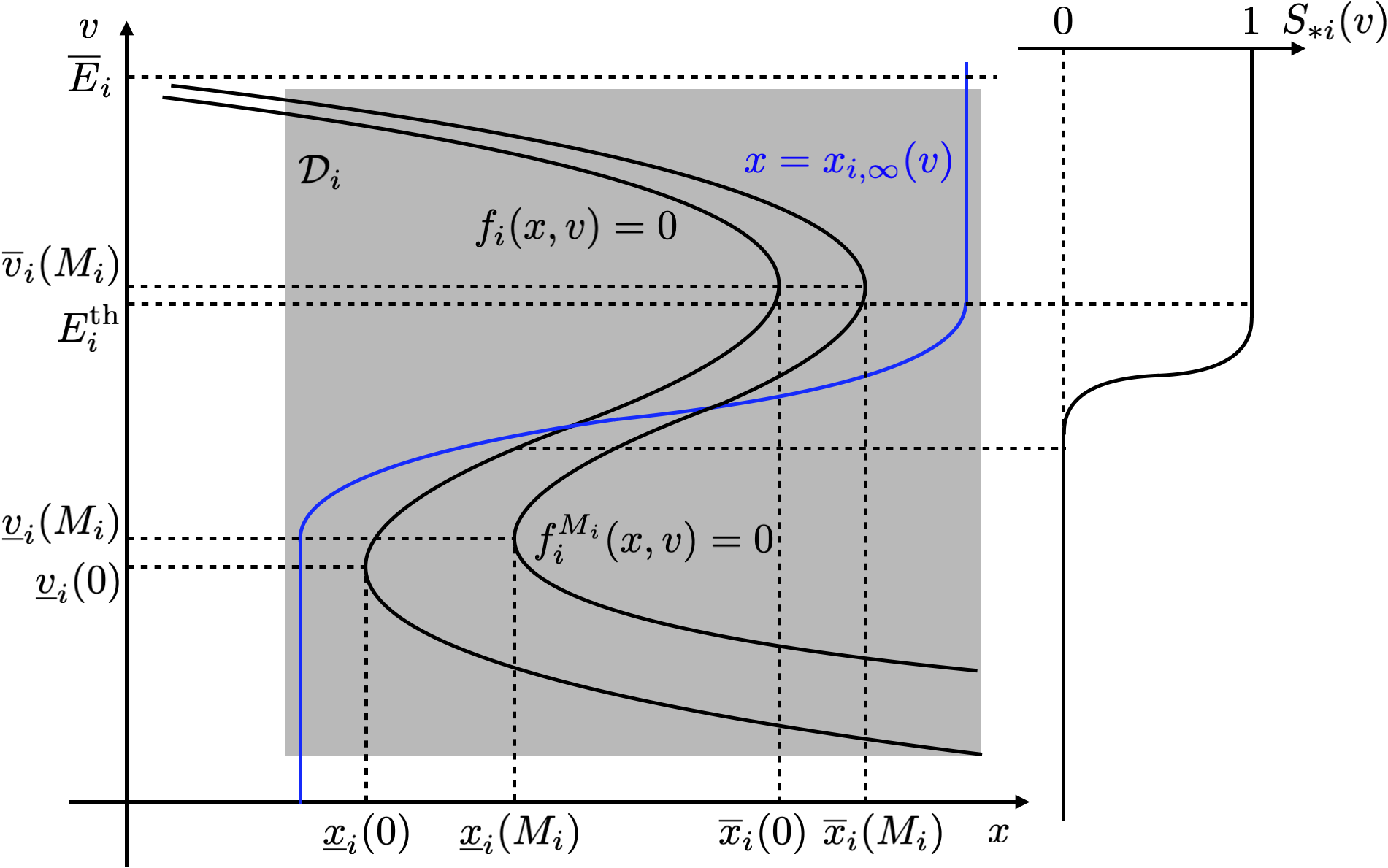}
\caption{An illustration of Assumptions~\ref{assum:spike}--\ref{assum:robust} and~\ref{assum:local} (see text for notations).}
\label{fig:phase}
\end{center}
\end{figure}

\begin{assumE}[Spiking phase portrait]\label{assum:spike}
The nullcline $x = x_{i, \infty}(v)$ has a monotonic shape and the nullcline $f_i(x, v) = 0$ is $N$-shaped, as illustrated in Figure~\ref{fig:phase}.
The nullcline $x = x_{i, \infty}(v)$ is on the left (right) to the lower (upper) branch of the nullcline $f_i(x, v) = 0$, hence there is only one intersection between the two, which is located in the middle.
We have $f_i(x, v) > 0$ ($<0$) in the left (right) region of the nullcline $f_i(x, v) = 0$.
\QEDB
\end{assumE}

Under Assumption~\ref{assum:spike}, the Poincare-Bendixson theorem ensures that all solutions (except the single unstable equilibrium) converge to a limit cycle.
One easily determines a positively invariant compact set $\mathcal{D}_i$ that includes the limit cycle (the grey box in Figure~\ref{fig:phase}).
The external input $(\overline{E}_i - v_i)G_i$ is called `excitatory,' as the input gain $(\overline{E}_i - v_i)$ is positive inside the positively invariant set $\mathcal{D}_i$.

The spiking nature of the limit cycle is ensured by the additional assumption that the time constant $\tau_{i, \epsilon}(\cdot)$ is $O(1)$ in the lower branch, while $O(\epsilon^q)$ with some $q \in (0, 1)$ in the upper branch of the nullcline $f_i(x, v) = 0$.
Indeed, as $\epsilon > 0$ becomes small, this ensures that the limit cycle exhibits a slow-fast-ultrafast behavior of spiking: the slow part of the limit cycle oscillation is a slow slide along the slow (lower) branch, whereas the ultrafast part is a switch between the lower branch and the upper branch, and finally the fast part is a fast slide along the fast (upper) branch.
Such assumptions are standard in biophysical models of neuronal models: significant time scale separation between the voltage dynamics and the ion channel kinetics is only pronounced in the subthreshold region.

\begin{assumE}[Biophysical nature of spiking]\label{assum:bio}
We assume that 
$$\bar{\tau}_i(v) := \lim_{\epsilon \to 0} \tau_{i, \epsilon}(v) \ge 0 \,\,\,  \text{ and } \,\,\,\bar{\lambda}_i(v) := \lim_{\epsilon \to 0} \frac{\epsilon^q}{\tau_{i, \epsilon}(v)} < \infty$$
are well-defined in the set $\mathcal{D}_i$, and satisfy
\begin{align*}
\bar{\tau}_i(v) &\begin{cases} = 0, &\mbox{ if } v \ge E_i^\text{th} \\ 
> 0, &\mbox{ if } v < E_i^\text{th}\end{cases} \quad \text{ and }  \\
\bar{\lambda}_i(v) &\begin{cases} > 0, &\mbox{ if } v > E_i^\text{th} \\
= 0, &\mbox{ if } v \le E_i^\text{th} \end{cases}
\end{align*}
with some $E_i^\text{th}$ that is located between the upper branch and the lower branch, as illustrated in Figure~\ref{fig:phase}.
\QEDB
\end{assumE}

The spiking behavior described above for an isolated neuron ($G_i = 0$) is \emph{robust} to the perturbations induced by synaptic coupling.
This is formalized in the following assumption.

\begin{assumE}[Robustness margin]\label{assum:robust}
We assume that there exists $M_i > 0$, such that for any $m \in [0, M_i]$, the perturbed system of~\eqref{eq:ind_neu} represented as
\begin{align*}
\tau_{i, \epsilon}(v_i)\dot{x}_i &= -x_i + x_{i, \infty}(v_i), \\ 
\epsilon \dot{v}_i &= f_i(x_i, v_i) + (\overline{E}_i - v_i)m =: f_i^{m}(x_i, v_i),
\end{align*}
has the same $\mathcal{D}_i$ as its positively invariant set, and satisfies Assumptions~\ref{assum:spike} and~\ref{assum:bio} with the same $E_i^\text{th}$.
In particular, we assume $\max_{m \in [0, M_i]} \underline{v}_i(m) < E_i^\text{th} < \min_{m \in [0, M_i]} \overline{v}_i(m)$, where $(\underline{x}_i(m), \underline{v}_i(m))$ ($(\overline{x}_i(m), \overline{v}_i(m))$) denotes the left (right) knee (that is, the extreme left (right) point of the nullcline $f_i^m(x, v) = 0$) of the perturbed system. 
We also assume $\underline{x}_i(M_i) < \overline{x}_i(0)$, as illustrated in Figure~\ref{fig:phase}.
\QEDB
\end{assumE}

\begin{lem1}\label{lem:move_right}
Under Assumptions~\ref{assum:spike} and~\ref{assum:robust}, the nullcline $f_i^m(x, v) = 0$ moves to the right as $m$ increases, i.e., for each $v < \overline{E}_i$, we have $(\partial x_i^m/\partial m)(v) > 0$ for all $m \in (0, M_i)$, where $x_i^m(v)$ denotes the unique root of $f_i^m(x, v) = 0$.
In particular, the left knee, the right knee, and the intersection move to the right as $m$ increases.
\QEDB
\end{lem1}

\begin{pf}
Since Assumption~\ref{assum:spike} is satisfied for all $m \in [0, M_i]$, for each $v < \overline{E}_i$, there exists a unique root $x_i^m(v)$ of $f_i^m(x, v) = 0$.
Moreover, since $f_i^m(x, v) < 0$ ($>0$) for $x > x_i^m(v)$ ($<x_i^m(v)$), we have $(\partial f_i/\partial x)(x_i^m(v), v) = (\partial f_i^m/\partial x)(x_i^m(v), v) < 0$.
Now, this implies from
$$\frac{\partial f_i}{\partial x}(x_i^m(v), v)\frac{\partial x_i^m}{\partial m}(v) + (\overline{E}_i - v) = \frac{\partial }{\partial m} f_i^m(x_i^m(v), v) = 0$$
that $(\partial x_i^m/\partial m)(v) > 0$ for $v < \overline{E}_i$.
\QEDBB
\end{pf}

Assumption~\ref{assum:robust} also illustrates the nature of the excitatory coupling: a weak pulse $G_i$ creates an open set near the left knee of the isolated system~\eqref{eq:ind_neu}, {\it advancing} the timing of the next spike.

\subsection{Synaptic coupling}\label{subsec:syn}

A standard model of excitatory synaptic coupling has the form
\begin{align}\label{eq:syn}
G_i(t) = g_i\sum_{j \in \mathcal{N}_i}\alpha_{ij}S_{ij}(v_j(t))
\end{align}
where $\mathcal{N}_i$ is the index set of presynaptic neurons, $\alpha_{ij} >0$ is the adjacency weight, and $S_{ij}(\cdot) : \mathbb{R} \to [0, 1]$ is the coupling function of sigmoidal type ($S_{ij}(\cdot)$ is non-decreasing).
The constant $g_i$ controls the coupling strength:

\begin{assumW}[Weak coupling]\label{assum:weak}
The constant $g_i > 0$ is chosen small enough to satisfy
$$g_i\sum_{j \in \mathcal{N}_i}\alpha_{ij} =: g_id_i \le M_i,$$
where $M_i$ is the robustness margin in Assumption~\ref{assum:robust}.
\QEDB
\end{assumW}

Our next assumption highlights the essential characteristics of excitatory synaptic coupling in the rapid and robust synchronization phenomenon: the coupling gain (determined by the derivative of the synaptic current) is high only near the threshold.

\begin{assumW}[Localized coupling]\label{assum:local}
For $j \in \mathcal{N}_i$,
$$S_{ij}(v) = \begin{cases} 1, \mbox{ if } v \ge E_j^\text{th}, \\
0, \mbox{ if } v \le \max_{m \in [0, M_j]} \underline{v}_j(m),
\end{cases}$$
where $\underline{v}_j(m)$ is defined in Assumption~\ref{assum:robust}.
\QEDB
\end{assumW}

As illustrated in the Introduction, under Assumptions~\ref{assum:spike}--\ref{assum:robust} and~\ref{assum:weak}--\ref{assum:local}, rapid convergence of each neuron to a neighborhood of its limit cycle attractor is guaranteed by treating the synaptic coupling as a weak exogenous disturbance (the first step of the convergence): every trajectory converges to $\mathcal{D}:=\mathcal{D}_1 \times \cdots \times \mathcal{D}_N$.
Hence, individual neurons persistently spike regardless of the coupling.
Next, we turn our attention to the network behavior of those spiking oscillators.

Recall that any directed graph $\mathcal{G}$ consists of multiple iSCCs and their followers.
The behavior of each iSCC is independent of the other iSCCs, due to the local interaction described in~\eqref{eq:syn}.
Therefore, we focus our attention on a single iSCC.
In particular, we assume the following.
The behavior of the followers corresponds to the phenomenon of entrainment, rather than synchronization.

\begin{assumG}[Connectivity]\label{assum:connec}
The communication graph $\mathcal{G} =(\mathcal{N}, \mathcal{E})$ determined by the neighborhoods $\mathcal{N}_i$ for $i \in \mathcal{N}$ is strongly connected.
\QEDB
\end{assumG}

\subsection{A network-wise phase reduction}\label{subsec:fs_net}

The analysis of the network relies on its three-time scale structure.
Our biophysical assumptions (Assumptions~\ref{assum:spike}--\ref{assum:robust} and~\ref{assum:weak}--\ref{assum:local}) illustrated in the previous subsections ensure that, for the $\epsilon = 0$ limit, a trajectory that starts from the slow region
\begin{align}\label{eq:region_slow}
\begin{split}
\mathcal{S} &:= \left\{{\rm col}(x_1, v_1^\text{sm}(x_1), \dots, x_N, v_N^\text{sm}(x_N)) : \right. \\
&\,\,\,\quad\quad\quad\quad\quad\quad\quad\quad \left. x_i \in [\underline{x}_i(0), \overline{x}_i(M_i)], i \in \mathcal{N}\right\},
\end{split}
\end{align}
where $v_i^\text{sm}(x)$ denotes the smallest root of $f_i(x, v) = 0$ (i.e., $\min \{v : x_i^0(v) = x\}$), returns back to the slow region $\mathcal{S}$ after an ultrafast-fast transition of spiking.

Note that the network is completely decoupled in the slow region $\mathcal{S}$:
\begin{align}\label{eq:net_slow}
\begin{split}
\dot{x}_i &= \frac{-x_i + x_{i, \infty}(v_i^\text{sm}(x_i))}{\bar{\tau}_i(v_i^\text{sm}(x_i))} =: h_i(x_i), \\
v_i &= v_i^\text{sm}( x_i), \quad\quad\quad\quad\quad\quad\quad\quad\quad\quad i \in \mathcal{N},
\end{split}
\end{align}
each neuron independently travels to the left along the lower branch according to~\eqref{eq:net_slow}, until one of them reaches the left knee $(\underline{x}_i(0), \underline{v}_i(0))$.
The set of all such endpoints is 
\begin{align}\label{eq:large_Poincare}
\begin{split}
\mathcal{P}^\text{sp} &:= \left\{{\rm col}(x_1, v_1, \dots, x_N, v_N) \in \mathcal{S} : \exists i \text{ s.t. } x_i = \underline{x}_i(0)\right\}.
\end{split}
\end{align}
Any point in the set $\mathcal{P}^\text{sp}$ is a jump point where the flow must be continued as a solution of the ultrafast-fast behavior.
Assumptions~\ref{assum:spike}--\ref{assum:robust} and~\ref{assum:weak}--\ref{assum:local} ensure that the ultrafast-fast behavior that starts from a jump point in $\mathcal{P}^\text{sp}$ eventually returns to the region $\mathcal{S}$, and hence further back to another jump point of $\mathcal{P}^\text{sp}$ via another round of slow travel.
The spiking map from $\mathcal{P}^\text{sp}$ to $\mathcal{S}$ is denoted by
\begin{align}\label{eq:spike_map}
X^+(x_1, \dots, x_N) := {\rm col}(x_1^+, \dots, x_N^+) : \mathcal{P}^\text{sp} \to \mathcal{S}.
\end{align}
Note that the mapping $x_i^+(x_1, \dots, x_N)$ only uses $x_j$, $j \in \mathcal{N}$ as their argument for simplicity of notation, as $v_i$ is entirely determined by the relation $v_i^\text{sm}(x_i)$, in the slow region $\mathcal{S}$.

Singular perturbation analysis (see Appendix~\ref{app:just} for details of the argument) rests on the slow behavior in the singular limit $\epsilon = 0$.
This behavior is a pulse-coupled integrate-and-fire model of the type studied in~\citep{mirollo1990synchronization}: the neurons follow the decoupled integrating flow~\eqref{eq:net_slow} between instantaneous spikes that define the spiking times.
This singular behavior is studied in the next section.

\section{Integrate-and-fire model}\label{sec:i_n_f}

The integrate-and-fire model is represented as
\begin{align}\label{eq:i_n_f}
\begin{split}
\dot{x}_i &= h_i(x_i) < 0, \,\,\,\quad\quad\quad\quad\quad\quad x_i \in [\underline{x}_i, \overline{x}_i], \\
x_i^+ &= x_i^+(x_1, \dots, x_N) \in [\underline{x}_i, \overline{x}_i], \,\,\,\, \exists j\text{ s.t. } x_j = \underline{x}_j, \\
&\,\,\quad\quad\quad\quad\quad\quad\quad\quad\quad\quad\quad i \in \mathcal{N} := \{1, \dots, N\}.
\end{split}
\end{align}

The state-space of this behavior is the set $\mathcal{S}$:
$$\mathcal{S} := \{{\rm col}(x_1, \dots, x_N): x_i \in [\underline{x}_i, \overline{x}_i], i \in \mathcal{N}\}.$$
The integrate-and-fire model has the following properties with $\underline{x}_i^\text{mid}, \overline{x}_i^\text{mid} \in (\underline{x}_i, \overline{x}_i)$ such that $\underline{x}_i^\text{mid} < \overline{x}_i^\text{mid}$, $i \in \mathcal{N}$.

This integrate-and-fire model corresponds to the singular limit $\epsilon = 0$ of the network dynamics.
The goal of the two following sections is to establish that this model possesses a hyperbolic fixed point that corresponds to synchronous firing.

\begin{property}[A network-wise lower threshold]\label{property:low_thres}
When neuron $j \in \mathcal{N}$ reaches a lower threshold $\underline{x}_j$, the network spikes: there exists a nonempty index set $\mathcal{N}^\mathcal{F}(x_1, \dots, x_N) \subseteq \mathcal{N}$ of synchronously firing neurons. 
In particular, $j \in \mathcal{N}^\mathcal{F}(x_1, \dots, x_N)$.
\QEDB
\end{property}

Let us denote the set of all points where a network-wise spiking happens (uniformly) as $\mathcal{P}^\text{sp}$:
$$\mathcal{P}^\text{sp} := \{{\rm col}(x_1, \dots, x_N) \in \mathcal{S} : \exists j \text{ s.t. } x_j = \underline{x}_j\}.$$
Then, the return map is well-defined as
$$X^{++}(X^+(p)) : \mathcal{P}^\text{sp} \to \mathcal{P}^\text{sp},$$
where $X^{++}(p^+) := {\rm col}(x_1^{++}(p^+), \cdots, x_N^{++}(p^+))$,
\begin{align*}
x_i^{++}(x_1^+, \dots, x_N^+) &:= \tau_i^{-1}\left(\tau_i(x_i^+) - \min_{j\in\mathcal{N}} \tau_j(x_j^+)\right), \\
\tau_i(x_i) &:= \int_{x_i}^{\underline{x}_i} \frac{1}{h_i(x)} dx \ge 0, \quad x_i \in [\underline{x}_i, \overline{x}_i], 
\end{align*}
for $i \in \mathcal{N}$, and $X^+(p) := {\rm col}(x_1^+(p), \dots, x_N^+(p))$.
Note that $\tau_i(x_i)$ represents the time of travel of neuron $i$ from $x_i$ to the lower threshold $\underline{x}_i$.

\begin{property}[Local interaction]\label{property:local_net}
Neuron $i$ with $x_i > \underline{x}_i$ spikes only when a presynaptic neuron spikes and $x_i \le \underline{x}_i^\text{mid}$.
In particular, $i \in \mathcal{N}^\mathcal{F}(x_1, \dots, x_N)$ only if there exists a set of neurons $i_0, \dots, i_{l-1} \in \mathcal{N}^\mathcal{F}(x_1, \dots, x_N)$ such that $x_{i_0} = \underline{x}_{i_0}$ and $i_{k-1} \in \mathcal{N}_{i_{k}}$ for all $k = 1, \dots, l$ with $i_l=i$ (chain reaction), where $\mathcal{N}_i \subset \mathcal{N}$ is the index set of presynaptic neurons.

On the other hand, whenever a presynaptic neuron spikes, neuron $i$ with $x_i > \underline{x}_i$ synchronously spikes provided that it is sufficiently near the threshold $\underline{x}_i$.
\QEDB
\end{property}

We define the set of network synchrony as 
$$\mathcal{P}^\text{syn} := \{p \in \mathcal{P}^\text{sp} : \mathcal{N}^\mathcal{F}(p) = \mathcal{N}\}.$$
The connectivity of the graph determined by the neighborhoods $\mathcal{N}_i$ for $i \in \mathcal{N}$ (Assumption~\ref{assum:connec}) ensures by Property~\ref{property:local_net} that for each $j \in \mathcal{N}$, there exists a point ${\rm col}(x_1, \dots, x_N)$ inside the closure $\overline{\mathcal{P}^\text{syn}}$ of $\mathcal{P}^\text{syn}$ such that $x_j = \underline{x}_j$ and $x_i > \underline{x}_i$ for all $i \neq j$.

\begin{property}[Convexity]\label{property:convex}
For any two points $p = {\rm col}(x_1, \dots, x_N), p' = {\rm col}(x_1', \dots, x_N')\in \mathcal{P}^\text{sp}$ such that
$$x_i \le x_i', \quad \forall i \in \mathcal{N},$$
we have
$$\mathcal{N}^\mathcal{F}(x_1, \dots, x_N) \supseteq \mathcal{N}^\mathcal{F}(x_1', \dots, x_N').$$
In particular, for any monotonically increasing (i.e., at least one is strictly increasing) path that connects $p$ and $p'$, the points $p''$ where $\mathcal{N}^\mathcal{F}(p'')$ changes are isolated.
\QEDB
\end{property}

By Property~\ref{property:convex}, with a finite subset $\mathcal{B}$ of $\mathcal{B}^\text{syn}$, $\mathcal{P}^\text{syn}$ can be approximated as
\begin{align}\label{eq:approx_P_syn}
\begin{split}
&\bigcup_{p = {\rm col}(x_1, \dots, x_N) \in \mathcal{B}} \left\{\underline{x}_{j(p)} \equiv x_{j(p)}\right\} \times \prod_{i \neq j(p)} \left[\underline{x}_i, x_i\right) \\
\subseteq & \bigcup_{p = {\rm col}(x_1, \dots, x_N) \in \mathcal{B}^\text{syn}} \{\underline{x}_{j(p)}\} \times \prod_{i \neq j(p)} [\underline{x}_i, x_i) = \mathcal{P}^\text{syn},
\end{split}
\end{align}
where $\mathcal{B}^\text{syn}$ is the set of all points $p = {\rm col}(x_1, \dots, x_N)$ on the boundary of $\mathcal{P}^\text{syn}$ such that there exists an index $j(p)$ such that $x_{i} = \underline{x}_{i}$ if and only if $i = j(p)$.
The equality holds except for a measure zero set.
Note that for~\eqref{eq:approx_P_syn} to be a good approximation, for each $j \in \mathcal{N}$, there exists at least one point $p \in \mathcal{B}$ such that $j(p) = j$.

Finally, regarding the spiking map $x_i^+(\cdot, \dots, \cdot)$, $i \in \mathcal{N}$, we have the following properties.

\begin{property}[Localized coupling]\label{property:spike_map}
If neuron $i$ spikes, then 
$$x_i^+(x_1, \dots, x_N) \in [\overline{x}_i^\text{mid}, \overline{x}_i].$$
On the other hand, if neuron $i$ does not spike, then ($\eta \ge 0$)
$$x_i^+(x_1, \dots, x_N) \in [\max\{x_i - \eta, \underline{x}_i\}, x_i].$$ 
If, in addition, $\mathcal{N}_i \cap \mathcal{N}^\mathcal{F}(x_1, \dots, x_N) = \emptyset$, then
$$x_i^+(x_1, \dots, x_N) = x_i.$$
\QEDB
\end{property}

Property~\ref{property:spike_map} illustrates that the network dynamics is almost decoupled away from the synchronous spike times.

\begin{property}[Stable spiking]\label{property:stab_spik}
The spiking map is continuous and differentiable almost everywhere in each region where $\mathcal{N}^\mathcal{F}(x_1, \dots, x_N)$ is invariant.
A network-wise spiking is stable: there exist functions $\overline{H}_i: [\underline{x}_i, \overline{x}_i] \to \mathbb{R}_{> 0}$, $i \in \mathcal{N}$ and $\overline{r} \ge 1$ such that when the spiking map is differentiable, for any infinitesimal perturbation ${\rm col}(\delta x_1, \dots, \delta x_N)$ that preserves $\mathcal{N}^\mathcal{F}(x_1, \dots, x_N)$ and an index $j$ such that $x_j = \underline{x}_j$ (so to remain in $\mathcal{P}^\text{sp}$), we have
\begin{align}\label{eq:mon_in_mean}
\begin{split}
&\max_{i \in \mathcal{N}^\mathcal{F}(p)} \frac{1}{\overline{H}_i(x_i^+(p))}\delta x_i^+ - \min_{i \in \mathcal{N}^\mathcal{F}(p)} \frac{1}{\overline{H}_i(x_i^+(p))}\delta x_i^+ \\
&\quad \le \overline{r} \left[\max_{i \in \mathcal{N}^\mathcal{F}(p)} \frac{1}{\overline{H}_i(x_i)}\delta x_i - \min_{i \in \mathcal{N}^\mathcal{F}(p)}\frac{1}{\overline{H}_i(x_i)}\delta x_i\right]
\end{split}
\end{align}
and $\delta x_i^+ = \delta x_i$, $i \notin \mathcal{N}^\mathcal{F}(p)$, where $p := {\rm col}(x_1, \dots, x_N)$.

\QEDB
\end{property}

Properties~\ref{property:low_thres}--\ref{property:stab_spik} are the key properties enabling synchrony in earlier integrate-and-fire models of the literature such as those studied in~\citep{mirollo1990synchronization,mauroy2008clustering,mauroy2011dichotomic}.
In the rest of the section, we will show that they also hold for the spiking map $X^+(\cdot, \dots, \cdot)$ defined in~\eqref{eq:spike_map}, where $\underline{x}_i$, $\overline{x}_i$, $\underline{x}_i^\text{mid}$, and $\overline{x}_i^\text{mid}$ are $\underline{x}_i(0)$, $\overline{x}_i(M_i)$, $\underline{x}_i(M_i)$, and $\overline{x}_i(0)$, respectively.

\subsection{Absorption: synchronous spiking}\label{subsec:sync_jump}

Property~\ref{property:low_thres} has already been established in Section~\ref{subsec:fs_net}.
So, let us detail more on what happens after when the trajectory leaves the slow region $\mathcal{S}$ from $\mathcal{P}^\text{sp}$.
This corresponds to the ultrafast-fast behavior as explained in Section~\ref{subsec:fs_net}.
In detail, a trajectory that leaves from $\mathcal{P}^\text{sp}$ follows the ultrafast behavior 
\begin{align}\label{eq:net_fast}
\begin{split}
x_i' &= 0, \\
v_i' &= f_i(x_i, v_i) + g_i(\overline{E}_i - v_i)\sum_{j \in \mathcal{N}_i}  \alpha_{ij}S_{ij}(v_j), \,\,\,\,\, i \in \mathcal{N},
\end{split}
\end{align}
in the ultrafast region $\mathcal{UF} \subset \mathcal{D}\setminus\mathcal{S}$ (where the ultrafast time scale $\epsilon$ dominates), which is defined as the region where the vector field of the network~\eqref{eq:net_fast} is non-zero.

This corresponds to the phenomenon of {\it absorption}: firing neurons absorb neighboring neurons that are sufficiently near the threshold, resulting in synchronous firing~\citep{mirollo1990synchronization,mauroy2008clustering}.
More precisely, a set $\mathcal{N}^\text{rt}$ of neurons $i$ at their left knee $(\underline{x}_i(0), \underline{v}_i(0))$ spike, i.e., go over the threshold $E_i^\text{th}$ (because $v_i' > 0$ for all $v_i \in (\underline{v}_i(0), E_i^\text{th}]$), resulting in $S_{*i}(v_i) = 1$, which yields a shift of the nullcline of other neurons (neuron $j$ such that $i \in \mathcal{N}_j$) into the right, yielding another spike, all in the same ultrafast time scale.

In particular, for each synchronous spiking, there exists a unique subgraph $\mathcal{T} = (\mathcal{N}^\mathcal{T}, \mathcal{E}^\mathcal{T})$ of $\mathcal{G}$ with the hierarchical structure, such that a spike of the $0$-th layer $\mathcal{N}_0^\mathcal{T} = \mathcal{N}^\text{rt}$ (i.e., set of all roots) yields a spike of the next layer $\mathcal{N}_1^\mathcal{T}$, and so on to all the neurons in $\mathcal{N}^\mathcal{T} \subseteq \mathcal{N}$, where
$$\mathcal{N}_{l+1}^\mathcal{T} := \bigcup_{i \in \mathcal{N}_l^\mathcal{T}} \{j: i \in \mathcal{N}_{j, \mathcal{T}}\} \setminus \overline{\mathcal{N}}_l^\mathcal{T}, \quad l \ge 0,$$
$\mathcal{N}_{j, \mathcal{T}}$ is the set of neighbors associated with the subgraph $\mathcal{T}$, and $\overline{\mathcal{N}}_l^\mathcal{T} := \cup_{m=0}^l \mathcal{N}_m^\mathcal{T}$.
For point ${\rm col}(x_1, \dots, x_N) \in \mathcal{P}^\text{sp}$, the corresponding subgraph can be found as follows.
\begin{enumerate}[1)]
\item Let $l = 1$ and $\mathcal{N}_0^\mathcal{T} = \{j : x_j = \underline{x}_j(0)\}$.
\item Let $\mathcal{N}_l^\mathcal{T} = \{j : x_j < \underline{x}_j(g_j\overline{d}_j^\mathcal{T})\} \setminus \overline{\mathcal{N}}_{l-1}^\mathcal{T}$ and $\mathcal{E}_l^\mathcal{T} = \cup_{k \in \mathcal{N}_{l}^\mathcal{T}} ((\mathcal{N}_k \cap \mathcal{N}_{l-1}^\mathcal{T})\times \{k\})$, where $\overline{d}_j^\mathcal{T} := \sum_{k \in \mathcal{N}_j\cap\overline{\mathcal{N}}_{l-1}^\mathcal{T}} \alpha_{jk}$.
\item If $\mathcal{N}_l^\mathcal{T} \neq \emptyset$, then let $l = l+1$ and repeat 2).
Otherwise, let $l^* = l-1$ and define $\mathcal{T}$ as $(\mathcal{N}^\mathcal{T}, \mathcal{E}^\mathcal{T})$, where $\mathcal{N}^\mathcal{T} := \overline{\mathcal{N}}_{l^*}^\mathcal{T}$ and $\mathcal{E}^\mathcal{T} := \cup_{l=1}^{l^*} \mathcal{E}_l^\mathcal{T}$.
\end{enumerate}
Note that this means
\begin{align*}
x_j &\in X_j^{\mathcal{T}} := \begin{cases} \left[\underline{x}_j\left(g_j\underline{d}_j^\mathcal{T}\right), \underline{x}_j\left(g_j\overline{d}_j^\mathcal{T}\right)\right], &\mbox{ if } j \in \mathcal{N}_0^\mathcal{T}, \\
\left[\underline{x}_j\left(g_j\underline{d}_j^\mathcal{T}\right) , \underline{x}_j\left(g_j\overline{d}_j^\mathcal{T}\right)\right), &\mbox{ otherwise,}
\end{cases}
\end{align*}
where $\underline{d}_j^{\mathcal{T}} := \sum_{k \in \mathcal{N}_j \cap \overline{\mathcal{N}}_{l-1}^{\mathcal{T}}} \alpha_{jk}$ for neuron $j$ in the $(l+1)$-th layer and $\overline{\mathcal{N}}_{-1}^{\mathcal{T}} = \overline{\mathcal{N}}_{-2}^{\mathcal{T}} = \emptyset$.

\setcounter{propertyp}{1}
\begin{propertyp}[Fast threshold modulation]\label{propertyp:FTM}
Neuron~$i$ with $x_i > \underline{x}_i$ spikes if and only if there exists a subgraph $\mathcal{T} = (\mathcal{N}^\mathcal{T}, \mathcal{E}^\mathcal{T})$ of $\mathcal{G}$ such that $i \in \mathcal{N}^\mathcal{T}$ and 
$$x_k \in X_k^{\mathcal{T}} \subset [\underline{x}_k, \underline{x}_k^\text{mid}], \quad \forall k \in \mathcal{N}^\mathcal{T}.$$
In particular, for each $p \in \mathcal{P}^\text{sp}$, there exists such a subgraph $\mathcal{T}$ satisfying $\mathcal{N}^\mathcal{T} = \mathcal{N}^\mathcal{F}(p)$.
\QEDB
\end{propertyp}

This stronger property ensures that a collection $\mathcal{B}$ of explicit points indexed by a spanning tree $\mathcal{T}$ of $\mathcal{G}$ as
$$p^{\mathcal{B}_\mathcal{T}} := {\rm col}(\underline{x}_1(g_1\overline{d}_1^\mathcal{T}), \dots, \underline{x}_N(g_N\overline{d}_N^\mathcal{T})) \in \mathcal{B}^\text{syn},$$
exactly characterizes $\mathcal{P}^\text{syn}$, according to~\eqref{eq:approx_P_syn}.
In particular,
\begin{align}\label{eq:charac_P_syn}
\begin{split}
&\bigcup_{\mathcal{T} \text{ of } \mathcal{G}} \left\{\underline{x}_{j(p^{\mathcal{B}_\mathcal{T}})}\right\} \times \prod_{i \neq j(p^{\mathcal{B}_\mathcal{T}})} \left[\underline{x}_i, \underline{x}_i(g_i\overline{d}_i^\mathcal{T})\right) = \mathcal{P}^\text{syn}
\end{split}
\end{align}
holds except for a measure zero set.

We emphasize that the nodal excitability (as illustrated at the end of Section~\ref{subsec:spike}) and the weak localized synaptic coupling (as emphasized in Section~\ref{subsec:syn}) are essential to provide this open set of synchronous spiking: a strong contraction for synchronous spiking behavior, which provides both the rapid convergence and the robustness with respect to heterogeneity (see Sections~\ref{sec:main} and~\ref{sec:main_bio}).
This is akin to the mechanism of fast threshold modulation: the coupling acts mainly to create an open transitional region.

\begin{rem1}
If we linearize the ultrafast behavior~\eqref{eq:net_fast}, which corresponds to a synchronous spiking, then we observe diffusive coupling with a time-varying edge-wise coupling gain $k_{ij}(\cdot)$ that is temporarily strong:
\begin{align*}
\delta v_i' &= \frac{\partial f_i}{\partial x}(x_i(t), v_i(t))\delta x_i + a_i(t) \delta v_i \\
&\quad  + \sum_{j \in \mathcal{N}_i}k_{ij}(t) [\delta v_j - \delta v_i],
\end{align*}
where $k_{ij}(t) := g_i(\overline{E}_i - v_i(t)) \alpha_{ij} S_{ij}'(v_j(t))$ and $a_i(t) := \frac{\partial f_i}{\partial v}(x_i(t), v_i(t)) - g_i\sum_{j \in \mathcal{N}_i}\alpha_{ij}S_{ij}(v_j(t))+ \sum_{j \in \mathcal{N}_i}k_{ij}(t)$.
This strong pulling in the ultrafast time scale is what provides a synchronous spiking of the next layer.
\QEDB
\end{rem1}

Property~\ref{property:convex} follows from the construction of the subgraph $\mathcal{T}$ for point $p \in \mathcal{P}^\text{sp}$.
In particular, $\overline{\mathcal{N}}_l^\mathcal{T} \supseteq \overline{\mathcal{N}}_l^{\mathcal{T}'}$ for all $l \ge 0$, where the subgraph $\mathcal{T}'$ corresponds to the point $p' \in \mathcal{P}^\text{sp}$.
In particular, $\mathcal{N}^\mathcal{F}(x_1, \dots, x_N)$ changes only when $x_j = \underline{x}_j(g_j\underline{d}_j^\mathcal{T}) > \underline{x}_j(0)$ or $x_j = \underline{x}_j(g_j\overline{d}_j^\mathcal{T})$ for some $j\notin \mathcal{N}_0^{\mathcal{T}}$.

Now, since the ultrafast behavior~\eqref{eq:net_fast} preserves the value $x_i$ during its entire transition, a trajectory that starts from $\mathcal{P}^\text{sp}$ ends up (via ultrafast behavior) in the fast region $\mathcal{F}$ (that is $\mathcal{S} \to \mathcal{UF} \to \mathcal{F}$), which is the set of all points ${\rm col}(x_1, v_1, \dots, x_N, v_N) \in \mathcal{D} \setminus (\mathcal{S} \cup \mathcal{UF})$ such that there exists a nonempty index set $\mathcal{N}^\mathcal{F} \subseteq \mathcal{N}$ satisfying 
\begin{align*}
x_i &\in \begin{cases} [\underline{x}_i(0), \overline{x}_i(m_i)] \supset [\underline{x}_i(g_i\underline{d}_i^\mathcal{T}), \underline{x}_i(g_i\overline{d}_i^\mathcal{T})], &\mbox{ if } i \in \mathcal{N}^\mathcal{F}, \\
[\underline{x}_i(m_i), \overline{x}_i(M_i)], &\mbox{ if } i \notin \mathcal{N}^\mathcal{F},\end{cases} \\
v_i &= \begin{cases}v_i^\text{lg}(m_i, x_i), &\mbox{ if } i \in \mathcal{N}^\mathcal{F}, \\
v_i^\text{sm}(m_i, x_i), &\mbox{ if } i \notin \mathcal{N}^\mathcal{F},\end{cases}\\
m_i &= g_i\sum_{j \in \mathcal{N}_i \cap \mathcal{N}^\mathcal{F}}\alpha_{ij}, \quad\quad\quad\quad\quad\quad\quad\quad\quad\quad\quad\quad i \in \mathcal{N},
\end{align*}
where $v_i^\text{lg}(m, x)$ ($v_i^\text{sm}(m, x)$) denotes the largest (smallest) root of $f_i^m(x, v) = 0$ (i.e., $\max\{v : x_i^m(v) = x\}$ ($\min\{v : x_i^m(v) = x\}$)) for $x \le \overline{x}_i(m)$ ($x \ge \underline{x}_i(m)$).
Note that $v_i^\text{sm}(\cdot) = v_i^\text{sm}(0, \cdot)$.
In the next subsection, we will detail on the fast behavior in the fast region $\mathcal{F}$, where the fast time scale $\epsilon^q$ dominates.
In doing so, we will more detail on the exact spiking map $x_i^+(\cdot, \dots, \cdot)$, $i \in \mathcal{N}$ defined in~\eqref{eq:spike_map}, and its properties.

\subsection{Spiking map}\label{subsec:spik_map}

To establish Property~\ref{property:spike_map}, we first note that the network is again completely decoupled in the fast region $\mathcal{F}$:
\begin{align}\label{eq:net_nor}
\begin{split}
x_i' &= \bar{\lambda}_i(v_i^\text{lg}(m_i, x_i))\left[-x_i + x_{i, \infty}(v_i^\text{lg}(m_i, x_i))\right]  \\
&=: H_i^{m_i}(x_i), \quad (v_i = v_i^\text{lg}(m_i, x_i)), \,\,\,\,\,\,\quad i \in \mathcal{N}^\mathcal{F}, \\
x_i' &= 0, \,\,\,\,\,\,\,\,\,\quad\quad\quad (v_i = v_i^\text{sm}(m_i, x_i)), \,\,\,\,\quad i \notin \mathcal{N}^\mathcal{F},
\end{split}
\end{align}
each neuron $i \in \mathcal{N}^\mathcal{F}$ independently travels to the right along the upper branch according to~\eqref{eq:net_nor}, until one of them reaches the right knee $(\overline{x}_i(m_i), \overline{v}_i(m_i))$.
Any such point is again a jump point where the flow must be continued as a solution of the ultrafast behavior~\eqref{eq:net_fast}.

Similar to the mechanism of synchronous spiking illustrated in the previous subsection, this yields that a subset $\mathcal{N}^{\mathcal{F}'} \subseteq \mathcal{N}^\mathcal{F}$ of neurons jumps to their lower branch while preserving their $x_i$ value, hence $x_i \in [\overline{x}_i(0), \overline{x}_i(m_i)]$.
If there are still some neurons remaining in the upper branch, i.e., $\mathcal{N}^\mathcal{F} \setminus \mathcal{N}^{\mathcal{F}'} \neq \emptyset$, then it means that we again enter the fast region $\mathcal{F}$, but with a different index set $\mathcal{N}^\mathcal{F} \setminus \mathcal{N}^{\mathcal{F}'}$.
This will change $m_i$, hence neurons remaining in the upper branch have a small jump before it moves again along the fast upper branch according to~\eqref{eq:net_nor}.

After possibly a few of these iterations ($\mathcal{F} \to \mathcal{UF} \to \mathcal{F}$), all the neurons in the upper branch will eventually jump to their lower branch ($\mathcal{F} \to \mathcal{UF} \to \mathcal{S}$), which completes the mechanism of spiking.
Note that the $x_i$ value of the neurons that previously jumped to their lower branch is preserved.
The number of these iterations ($\mathcal{F} \to \mathcal{UF} \to \mathcal{F}$) is bounded by $|\mathcal{N}^\mathcal{F}| - 1 \le N-1$.

\setcounter{propertyp}{3}
\begin{propertyp}[Localized coupling]\label{propertyp:spike_map}
If neuron $i$ spikes, then
\begin{align*}
x_i^+(x_1, \dots, x_N) &\in [\overline{x}_i(0),\overline{x}_i(g_id_i)] \subset [\overline{x}_i^\text{mid}, \overline{x}_i].
\end{align*}
If neuron $i$ does not spike, then $x_i^+(x_1, \dots, x_N) = x_i$.
\QEDB
\end{propertyp}

Next, consider the propagation of an infinitesimal perturbation ${\rm col}(\delta x_1, \dots, \delta x_N)$ (that preserves $\mathcal{N}^\mathcal{F}(x_1, \dots, x_N)$ and an index $j$ such that $x_j = \underline{x}_j$) through one iteration of $\mathcal{F} \to \mathcal{UF} \to \mathcal{F} \text{ or } \mathcal{S}$.
For this purpose, assume that a subset $\mathcal{N}^{\mathcal{F}'} \subseteq \mathcal{N}^\mathcal{F}$ of neurons jumps to their lower branch after this iteration.
Note that this iteration is differentiable if $\mathcal{N}^{\mathcal{F}'}$ is preserved in an open neighborhood, and is continuous even if $\mathcal{N}^{\mathcal{F}'}$ changes.

Since the fast behavior just corresponds to the shift of time spent on the fast upper branch, our infinitesimal perturbation can be redefined as
$$\delta T_i := \delta x_i/H_i^{m_i}(x_i), \quad i \in \mathcal{N}^\mathcal{F}$$
and accordingly
$$\delta T_i^+ := \delta x_i^+/H_i^{m_i}(x_i^+), \quad i \in \mathcal{N}^\mathcal{F},$$
for simplicity of representation, where $x_i^+$ and $\delta x_i^+$ are the corresponding values after the iteration.
Then, we have
$$\delta T_k^+ - \delta T_k = \delta T_i^+ - \delta T_i, \quad \forall i, k \in \mathcal{N}^\mathcal{F},$$
because
$$\int_{x_k}^{x_k^+}\frac{1}{H_k^{m_k}(x)} dx = \int_{x_i}^{x_i^+}\frac{1}{H_i^{m_i}(x)} dx, \quad \forall i, k \in \mathcal{N}^\mathcal{F}.$$
Note that there exists an index $k \in \mathcal{N}^{\mathcal{F}'}$ such that $\delta x_k^+ = 0$, as $x_k^+ = \overline{x}_k(m_k)$ (neuron $k$ reaches the right knee), hence
$$\max_{i \in \mathcal{N}^\mathcal{F}} \delta T_i^+ - \min_{i \in \mathcal{N}^\mathcal{F}} \delta T_i^+ \le \max_{i \in \mathcal{N}^\mathcal{F}} \delta T_i - \min_{i \in \mathcal{N}^\mathcal{F}} \delta T_i.$$

Note that between two iterations, we have to multiply
$$H_i^{m_i}(x_i^+)/H_i^{m_i^+}(x_i^+)$$
to $\delta T_i^+$ for neurons remaining in the upper branch, $i \in \mathcal{N}^\mathcal{F}\setminus\mathcal{N}^{\mathcal{F}'}$, as they experience a small jump before flowing again along the fast upper branch.
This establishes Property~\ref{property:stab_spik}.

\begin{propertyp}[Stable spiking]\label{propertyp:stab_spik}
Property~\ref{property:stab_spik} holds with 
\begin{align*}
\overline{r} &:= \left(\overline{r}_{\mathcal{F} \to \mathcal{F}}\right)^{N + 1} \ge 1, \\
\overline{r}_{\mathcal{F} \to \mathcal{F}} &:= \max_{i \in \mathcal{N}} \max_{m, m' \in [0, M_i]}\max_{x \in [\underline{x}_i, \overline{x}_i(\min\{m, m'\})]} \frac{H_i^m(x)}{H_i^{m'}(x)},
\end{align*}
and $\overline{H}_i(\cdot) := H_i^{M_i}(\cdot)$.
\QEDB
\end{propertyp}

Almost any trajectory that starts from the rest of the region $\mathcal{D} \setminus (\mathcal{S} \cup \mathcal{UF} \cup \mathcal{F})$ moves to $\mathcal{S}$ in a finite time, hence never going back to the rest of the region (see Appendix~\ref{app:almost_global}).

\begin{rem1}
This subsection shows the importance of a network-wise reduction for nonstandard singularly perturbed systems (nonstandard as in~\citep{wechselberger2020geometric}) that represent networks of neurons, instead of a neuron-wise reduction.
In particular, all the important interaction via weak localized synaptic coupling takes place in the ultrafast-fast time scale, making it difficult to characterize the network-wise spiking map $X^+(\cdot)$ via phase response curves.
\QEDB
\end{rem1}

\section{Main result: integrate-and-fire model}\label{sec:main}

It is known that synchronous spiking is not necessarily persistent in the presence of heterogeneity and a general network topology~\citep{mirollo1990synchronization,mauroy2008clustering}.
A synchronous spike of the entire network does not necessarily lead to a synchronous spike of the entire network after a return.

The goal of the present section is to detail the conditions that make perfect network synchrony stable and robust.

\subsection{Condition for the local stability of network synchrony}\label{subsec:local}

Assume for a moment that there exists a fixed point $p^*$ of the return map, i.e., $X^{++}(X^+(p^*)) = p^*$, in the interior of $\mathcal{P}^\text{syn}$ so that a synchronous spiking that starts from $p^*$ returns back to $p^*$.
Then, the local stability of the fixed point can be guaranteed by a contraction property, that is, the existence of $c_{\mathcal{S}\to \mathcal{F}} > 0$ and $c_{\mathcal{F} \to \mathcal{S}} > 0$ such that 
\begin{align}\label{eq:mon_in_mean_actual}
\begin{split}
|h_i(x_i^*)| &\le c_{\mathcal{S} \to \mathcal{F}} \overline{H}_i(x_i^*), \\
\overline{H}_i(x_i^+(p^*)) &\le c_{\mathcal{F} \to \mathcal{S}} |h_i(x_i^+(p^*))|, \quad \forall i \in \mathcal{N},\\
c_{\mathcal{S} \to \mathcal{F}}  c_{\mathcal{F}\to\mathcal{S}} &< 1/\overline{r} \le 1,
\end{split}
\end{align} 
where ${\rm col}(x_1^*, \dots, x_N^*) := p^* \in \mathcal{P}^\text{syn}$, and the constant $\overline{r} \ge 1$ and the function $\overline{H}_i(\cdot)$ are given in Property~\ref{property:stab_spik}.
Given that $x_i^* \le \underline{x}_i^\text{mid} < \overline{x}_i^\text{mid} \le x_i^+(p^*)$, this is merely equivalent to $|h_i(\cdot)|/\overline{H}_i(\cdot)$ being monotone increasing in the mean:
$$\frac{|h_i(x_i^+(p^*))|}{\overline{H}_i(x_i^+(p^*))} - \frac{|h_i(x_i^*)|}{\overline{H}_i(x_i^*)} \ge \frac{1}{c_{\mathcal{F} \to \mathcal{S}}} - c_{\mathcal{S} \to \mathcal{F}}> 0.$$

\begin{thm1}[Local stability]\label{thm:local_stab_i_n_f_fixed}
Given the integrate-and-fire model~\eqref{eq:i_n_f} with Properties~\ref{property:low_thres}--\ref{property:stab_spik}, in addition to Assumption~\ref{assum:connec} assume that there is a fixed point $p^*$ of the return map $X^{++}(X^+(\cdot))$ in the interior of $\mathcal{P}^\text{syn}$ that satisfies~\eqref{eq:mon_in_mean_actual} with some $c_{\mathcal{S}\to\mathcal{F}},c_{\mathcal{F}\to\mathcal{S}}$. 
Then, it is locally exponentially stable with the convergence rate $c_{\mathcal{S}\to\mathcal{F}}  \overline{r}  c_{\mathcal{F}\to\mathcal{S}} <1$.
\QEDB
\end{thm1}

\begin{rem1}
For this fixed point to have an open neighborhood of persistent synchronous spiking, there must exist a point $p \in \mathcal{P}^\text{syn}$ such that $x_i > \underline{x}_i$ for all neurons except one say $1$, i.e., $x_1 = \underline{x}_1$.
Then, by Property~\ref{property:local_net}, this implies that there exists a spanning tree having a root $1$.
So, as illustrated in Section~\ref{subsec:syn}, we need connectivity of $\mathcal{G}$.
\QEDB
\end{rem1}

\begin{pf}
Since there is an open neighborhood of the fixed point $p^* =: {\rm col}(x_1^*, \dots, x_N^*)$, its local exponential stability can be determined by the linearization of the return map.
In particular, since by Property~\ref{property:stab_spik}, the spiking map is differentiable almost everywhere in $\mathcal{P}^\text{syn}$, and for any infinitesimal perturbation ${\rm col}(\delta x_1, \dots, \delta x_N)$ that preserves $\mathcal{N}^\mathcal{F}(p^*) = \mathcal{N}$ and an index $j$ such that $x_j = \underline{x}_j$ (hence $\delta x_j = 0$), we have~\eqref{eq:mon_in_mean} with $\mathcal{N}^\mathcal{F} = \mathcal{N}$ and $p = p^*$.

Now, analogously as in Section~\ref{subsec:spik_map}, $X^{++}(\cdot)$ just corresponds to the shift of time spent, hence let us redefine our infinitesimal perturbation as
$$\delta \tau_i^+ := \frac{1}{h_i(x_i^+(p^*))} \delta x_i^+$$
and accordingly
$$\delta \tau_i^{++} := \frac{1}{h_i(x_i^{++}(X^+(p^*)))} \delta x_i^{++} \,\,\,\,\text{and}\,\,\,\, \delta\tau_i := \frac{1}{h_i(x_i^*)}\delta x_i,$$
for simplicity of representation.
Then, we have
$$\delta \tau_k^{++} - \delta \tau_k^+ = \delta \tau_i^{++} - \delta \tau_i^+, \quad \forall i, k \in \mathcal{N},$$
because
$$\int_{x_k^+}^{x_k^{++}} \frac{1}{h_k(x)} dx = \int_{x_i^+}^{x_i^{++}} \frac{1}{h_i(x)} dx, \quad \forall i, k \in \mathcal{N}.$$
Since $p^*$ is a fixed point, the infinitesimal perturbation ${\rm col}(\delta x_1^+, \dots, \delta x_N^+)$ again preserves the index $j$ such that $x_j^{++}(X^+(p^*)) = x_j^*= \underline{x}_j$, hence $\delta\tau_j^{++} = \delta x_j^{++} = 0$, hence
$$\max_{i \in \mathcal{N}} \delta \tau_i^{++} - \min_{i \in \mathcal{N}} \delta \tau_i^{++} \le \max_{i \in \mathcal{N}} \delta \tau_i^+ - \min_{i \in \mathcal{N}} \tau_i^+,$$
and thus, by the stability of the spiking (Property~\ref{property:stab_spik}),
\begin{align}\label{eq:metric_contract}
&\max_{i \in \mathcal{N}} \delta \tau_i^{++} - \min_{i \in \mathcal{N}} \delta \tau_i^{++} \nonumber \\
&\le 
c_{\mathcal{F} \to \mathcal{S}}  \left[\max_{i \in \mathcal{N}} \frac{1}{\overline{H}_i(x_i^+(p^*))} \delta x_i^+ - \min_{i \in \mathcal{N}} \frac{1}{\overline{H}_i(x_i^+(p^*))} \delta x_i^+\right] \nonumber\\
&\le c_{\mathcal{F}\to\mathcal{S}}  \overline{r} \left[\max_{i \in \mathcal{N}} \frac{1}{\overline{H}_i(x_i^*)}\delta x_i - \min_{i \in \mathcal{N}} \frac{1}{\overline{H}_i(x_i^*)} \delta x_i\right] \nonumber \\
&\le c_{\mathcal{F}\to\mathcal{S}}  \overline{r}  c_{\mathcal{S}\to\mathcal{F}}  \left[\max_{i \in \mathcal{N}} \delta\tau_i - \min_{i \in \mathcal{N}} \delta \tau_i\right].
\end{align}
This concludes the proof.
\QEDBB
\end{pf}

The local stability of a fixed point $p^*$, which is in the interior of $\mathcal{P}^\text{syn}$, ensures the existence of a neighborhood $\mathcal{P}^\text{pos}$ which is stable, that is $\mathcal{P}^\text{pos}$ is a positively invariant set in $\mathcal{P}^\text{syn}$: any synchronous spiking that starts from $\mathcal{P}^\text{pos}$ returns back to $\mathcal{P}^\text{pos}$.
In particular, there exists $\mathcal{P}^\text{pos} \subseteq \mathcal{P}^\text{syn}$, $c_{\mathcal{S}\to\mathcal{F}} > 0$, and $c_{\mathcal{F}\to\mathcal{S}} > 0$ that satisfies~\eqref{eq:mon_in_mean_actual} for all $p \in \mathcal{P}^\text{pos}$ instead of just $p^*$.
Note that by Properties~\ref{property:local_net} and~\ref{property:spike_map}, the region that corresponds to $x_i$ and the region that corresponds to $x_i^+(p)$ are separated, hence~\eqref{eq:mon_in_mean_actual} is well justified.
In particular, we have 
$$\underline{x}_i \le x_i \le \underline{x}_i^\text{mid} < \overline{x}_i^\text{mid} \le x_i^+(x_1, \dots, x_N) \le \overline{x}_i,$$
for any ${\rm col}(x_1, \dots, x_N) \in \mathcal{P}^\text{syn}$.
In this sense,~\eqref{eq:mon_in_mean_actual} for all $p \in \mathcal{P}^\text{pos}$ can be guaranteed by the following assumption.

\begin{assumM}[Monotonicity in the mean]\label{assum:mon_in_mean}
There exists $c_{\mathcal{S}\to\mathcal{F}}, c_{\mathcal{F}\to\mathcal{S}} > 0$ such that
\begin{align*}
\max_{x \in [\underline{x}_i, \underline{x}_i^\text{mid}]} \frac{|h_i(x)|}{\overline{H}_i(x)} &\le c_{\mathcal{S}\to\mathcal{F}}, \\
\max_{x \in [\overline{x}_i^\text{mid}, \overline{x}_i]} \frac{\overline{H}_i(x)}{|h_i(x)|} &\le c_{\mathcal{F}\to\mathcal{S}}, \quad\forall i \in \mathcal{N},
\end{align*}
and $c_{\mathcal{S}\to\mathcal{F}} c_{\mathcal{F}\to\mathcal{S}} < 1/\overline{r}$, where the constant $\overline{r} \ge 1$ and the function $\overline{H}_i(\cdot)$ are given in Property~\ref{property:stab_spik}.
\QEDB
\end{assumM}

A simple sufficient condition for Assumption~\ref{assum:mon_in_mean} is the positivity of the derivative: that there exists a sufficiently large $\lambda > 0$ such that 
$$\frac{d}{dx} \frac{|h_i(x)|}{\overline{H}_i(x)} \ge \lambda > 0, \quad \forall x \in (\underline{x}_i, \overline{x}_i), \quad \forall i \in \mathcal{N}.$$

Under this stability condition (Assumption~\ref{assum:mon_in_mean}), in particular, we can show contraction inside $\mathcal{P}^\text{syn}$: the contraction of the metric defined for any two points $p = {\rm col}(x_1, \dots, x_N)$ and $p' = {\rm col}(x_1', \dots, x_N')$ in $\mathcal{P}^\text{sp}$:
$$d(p, p') := \max_{i \in \mathcal{N}} [\tau_i(x_i)- \tau_i(x_i')] - \min_{i \in \mathcal{N}} [\tau_i(x_i) - \tau_i(x_i')].$$
As in~\citep{somers1993rapid}, the notion of phase difference depends on time difference along the trajectory, not differences in position in phase space.

\begin{thm1}[Contraction inside $\mathcal{P}^\text{syn}$]\label{thm:local_stab_i_n_f}
Given the integrate-and-fire model~\eqref{eq:i_n_f} with Properties~\ref{property:low_thres}--\ref{property:stab_spik}, let Assumptions~\ref{assum:connec} and~\ref{assum:mon_in_mean} holds. 
Then, we have
\begin{align}\label{eq:metric_contract_detail}
\begin{split}
&d(X^{++}(X^+(p)), X^{++}(X^+(p'))) \le c_{\mathcal{S}\to\mathcal{F}}  \overline{r}  c_{\mathcal{F}\to\mathcal{S}} d(p, p'),
\end{split}
\end{align}
for all $p, p' \in \mathcal{P}^\text{syn}$.
\QEDB
\end{thm1}

\begin{pf}
From the argument in the proof of Theorem~\ref{thm:local_stab_i_n_f_fixed}, we can analogously conclude~\eqref{eq:metric_contract} for any point along the path that connects $p$ and $p'$ inside $\mathcal{P}^\text{syn}$ for any eligible infinitesimal perturbation ${\rm col}(\delta x_1, \dots, \delta x_N)$.
Note that such a path exists, as $\mathcal{P}^\text{syn}$ is convex, from Property~\ref{property:convex}.
Then, by integrating this along the path, we can conclude~\eqref{eq:metric_contract_detail}.
In particular, we can construct a path $p(s) = {\rm col}(x_1(s), \dots, x_N(s)) : [0, 2] \to \mathcal{P}^\text{syn}$ as follows.
\begin{enumerate}
\item Let $p(0) = p$.
\item Let $\mathcal{N}' \subseteq\mathcal{N}$ be the set of all the indices~$i$ such that $\tau_i(x_i') \le \tau_i(x_i)$ and let $\overline{\tau} := \max_{i \in \mathcal{N}'} [\tau_i(x_i) - \tau_i(x_i')] \ge 0$.
Then, for each $s \in [0, 1]$, we define $p(s)$ as
\begin{align*}
x_i(s) &= \begin{cases} &x_i, \quad\quad\quad\quad\quad\quad \mbox{ if } i \notin \mathcal{N}', \\ 
&\tau_i^{-1}(\tau_i(x_i) - s\overline{\tau}), \\
&\quad \mbox{ if }  i \in \mathcal{N}' \text{ and } \tau_i(x_i) - \tau_i(x_i') > s\overline{\tau}, \\
&x_i', \\
&\quad\mbox{ if } i \in \mathcal{N}' \text{ and } \tau_i(x_i) - \tau_i(x_i') \le s\overline{\tau}.\end{cases}
\end{align*}
\item Finally, let $\underline{\tau} := - \min_{i \notin \mathcal{N}'} [\tau_i(x_i) - \tau_i(x_i')] \ge 0$, and for each $s \in [1, 2]$, define ${p}(s)$ as
\begin{align*}
{x}_i(s) &= \begin{cases} &x_i', \quad\quad\quad\quad\quad\quad\mbox{ if } i \in \mathcal{N}', \\
&\tau_i^{-1}(\tau_i(x_i) + (s-1)\underline{\tau}), \\
&\mbox{ if } i \notin \mathcal{N}' \text{ and } \tau_i(x_i') - \tau_i(x_i) > (s-1)\underline{\tau}, \\
&x_i', \\
&\mbox{ if } i \notin \mathcal{N}' \text{ and } \tau_i(x_i') - \tau_i(x_i) \le (s-1)\underline{\tau}. \end{cases}
\end{align*}
\end{enumerate}
By its construction, for $s \in [0, 1]$, $\delta\tau_i(s) \in \{0, \overline{\tau}\}$, $i \in \mathcal{N}$.
Similarly, we have for $s \in [1, 2]$, $\delta\tau_i(s) \in \{0, -\underline{\tau}\}$, $i \in \mathcal{N}$.
This now implies from~\eqref{eq:metric_contract} that
\begin{align*}
&\max_{i \in\mathcal{N}} \delta\tau_i^{++}(s) - \min_{i\in\mathcal{N}} \delta\tau_i^{++}(s) \\
&\,\,\,\quad\quad\quad\quad\quad\quad\quad\quad\quad\le \begin{cases} c_{\mathcal{S}\to\mathcal{F}}  \overline{r}  c_{\mathcal{F}\to\mathcal{S}} \overline{\tau}, &\mbox{ if } s \in [0, 1], \\ c_{\mathcal{S}\to\mathcal{F}}  \overline{r}  c_{\mathcal{F}\to\mathcal{S}} \underline{\tau}, &\mbox{ if } s\in [1, 2].\end{cases}
\end{align*}
Therefore, we can conclude that
\begin{align*}
&d(X^{++}(X^+(p)), X^{++}(X^+(p'))) \\
&= \max_{i \in\mathcal{N}} [\tau_i(x_i^{++}(X^+(p(0)))) - \tau_i({x}_i^{++}(X^+(p(2))))] \\
&\quad- \min_{i \in\mathcal{N}} [\tau_i(x_i^{++}(X^+(p(0)))) - \tau_i(x_i^{++}(X^+(p(2))))] \\
&\le \int_0^2  D^+  \max_{i \in\mathcal{N}} [\tau_i(x_i^{++}(X^+(p(0)))) - \tau_i({x}_i^{++}(X^+(p(s))))] ds \\
&\quad-  \int_0^2  D_+  \min_{i \in\mathcal{N}} [\tau_i(x_i^{++}(X^+(p(0))))  -  \tau_i({x}_i^{++}(X^+(p(s))))] ds \\
&\le \int_0^2 \max_{i \in\mathcal{N}} \delta\tau_i^{++}(s) - \min_{i\in\mathcal{N}} \delta\tau_i^{++}(s) ds \\
&\le c_{\mathcal{S}\to\mathcal{F}}  \overline{r}  c_{\mathcal{F}\to\mathcal{S}}[\overline{\tau} + \underline{\tau}] = c_{\mathcal{S}\to\mathcal{F}}  \overline{r}  c_{\mathcal{F}\to\mathcal{S}} d(p, p'),
\end{align*}
where $D^+$ ($D_+$) denotes the upper (lower) right-hand Dini derivative.
\QEDBB
\end{pf}

Theorem~\ref{thm:local_stab_i_n_f} removes the possibility of periodic solutions that reside inside $\mathcal{P}^\text{syn}$, and therefore, we will continuously focus our attention on a fixed point.
In this respect, Theorem~\ref{thm:local_stab_i_n_f} further implies that under Assumption~\ref{assum:mon_in_mean}, the existence of a fixed point in the interior of $\mathcal{P}^\text{syn}$ is equivalent to the existence of an open positively invariant set in $\mathcal{P}^\text{syn}$.
This positively invariant set becomes the domain of attraction due again to Theorem~\ref{thm:local_stab_i_n_f}.
An explicit sufficient condition for the existence of a fixed point in this context is given in the next subsection.

\subsection{Existence of a fixed point $p^*$ in the interior of $\mathcal{P}^\text{syn}$}\label{subsubsec:exist_P_per}

According to the illustration of the integrate-and-fire model~\eqref{eq:i_n_f} in Section~\ref{sec:i_n_f}, the only explicit knowledge that we have on the point after synchronous spiking of the entire network is that it is inside the set $\prod_{i \in \mathcal{N}} [\overline{x}_i^\text{mid}, \overline{x}_i]$.

Therefore, the only explicit sufficient condition for the existence of a fixed point $p^* \in \mathcal{P}^\text{syn}$, or equivalently, the existence of a positively invariant set $\mathcal{P}^\text{pos} \subseteq \mathcal{P}^\text{syn}$ that we can conclude based on Properties~\ref{property:low_thres}--\ref{property:stab_spik} should be $X^{++}(p) \in \mathcal{P}^\text{pos}$ for all $p \in \prod_{i \in \mathcal{N}} [\overline{x}_i^\text{mid}, \overline{x}_i]$ with some $\mathcal{P}^\text{pos} \subseteq \mathcal{P}^\text{syn}$.
Thus, the condition will be most relaxed for $\mathcal{P}^\text{pos} = \mathcal{P}^\text{syn}$.

In this regard, the most relaxed explicit sufficient condition for the existence of a fixed point $p^* \in \mathcal{P}^\text{syn}$ that we can conclude based on Properties~\ref{property:low_thres}--\ref{property:stab_spik} is
\begin{align}\label{eq:exp_suff_cond_prop}
\begin{split}
X^{++}(p) &\in \bigcup_{p' = {\rm col}(x_1', \dots, x_N') \in \mathcal{B}}  \{\underline{x}_{j(p')}\} \times \prod_{i \neq j(p')} [\underline{x}_i, x_i'), \\
&\,\,\,\,\quad\quad\quad\quad\quad\quad\quad\quad\quad \forall p \in \prod_{i \in \mathcal{N}} [\overline{x}_i^\text{mid}, \overline{x}_i],
\end{split}
\end{align}
which is based on the approximation of $\mathcal{P}^\text{syn}$ given in~\eqref{eq:approx_P_syn}.
In particular, under this sufficient condition, there exists a unique fixed point $p^*$ in the interior of $\mathcal{P}^\text{syn}$.

Now,~\eqref{eq:exp_suff_cond_prop} means that for each $p \in  \prod_{i \in \mathcal{N}} [\overline{x}_i^\text{mid}, \overline{x}_i]$, if $x_j^{++}(p) = \underline{x}_j$, then there exists a point $p' = {\rm col}(x_1', \dots, x_N') \in \mathcal{B}$ such that $j(p') = j$ and
$$\underline{x}_i \le x_i^{++}(p) < x_i', \quad \forall i \neq j.$$
In other words, $\tau_{j}(x_j) = \min_{i \in \mathcal{N}} \tau_i(x_i)$ and
\begin{align}\label{eq:basic}
0 = \tau_i(\underline{x}_i) \le \tau_i(x_i) - \tau_{j}(x_{j}) < \tau_i(x_i'), \quad \forall i \neq j,
\end{align}
where $p =: {\rm col}(x_1, \dots, x_N)$.
Now, when $x_{j}$ becomes smaller as $\overline{x}_{j}^\text{mid}$ and $x_i$ become larger as $\overline{x}_i$ for all $i \neq j$, we still have $x_{j}^{++}(p) = \underline{x}_{j}$, and this implies
\begin{align}\label{eq:exp_suff_time_cond}
\begin{split}
&\exists p'' = {\rm col}(x_1'', \dots, x_N'') \in \mathcal{B} \text{ s.t. } j(p'') = j \text{ and }\\
&\,\,\,\quad\quad\quad\quad\quad\tau_i(\overline{x}_i)  - \tau_i(x_i'') < \tau_{j}(\overline{x}_{j}^\text{mid}), \quad \forall i \neq j,
\end{split}
\end{align}
and $\tau_{j}(\overline{x}_{j}^\text{mid}) \le \tau_i(\overline{x}_i)$ for all $i \neq j$.
Therefore,~\eqref{eq:exp_suff_cond_prop} is equivalent to~\eqref{eq:exp_suff_time_cond} for all $j \in \mathcal{N}^\text{fast} \subseteq\mathcal{N}$, where
\begin{align}\label{eq:def_Cp}
\mathcal{N}^\text{fast} := \{j \in \mathcal{N} : \tau_j(\overline{x}_j^\text{mid}) \le \min_{i \in \mathcal{N}} \tau_i(\overline{x}_i)\},
\end{align}
because then for any $p = {\rm col}(x_1, \dots, x_N) \in \prod_{i \in \mathcal{N}} [\overline{x}_i^\text{mid}, \overline{x}_i]$ there exists $j \in \mathcal{N}^\text{fast}$ satisfying $x_j^{++}(p) = \underline{x}_j$, and thus,
\begin{align*}
\tau_i(x_i) - \tau_i(x_i'') &\le \tau_i(\overline{x}_i) - \tau_i(x_i'') \\
&< \tau_j(\overline{x}_j^\text{mid}) \le \tau_j(x_j) \le \tau_i(x_i), \quad \forall i \neq j.
\end{align*}
A simple sufficient condition is~\eqref{eq:exp_suff_time_cond} for all $j \in \mathcal{N}$.
In particular, under this sufficient condition, there exists a fixed point $p^*$ inside
\begin{align}\label{eq:def_P_pos}
\bigcup_{j \in \mathcal{N}} \{\underline{x}_j\} \times \prod_{i \neq j} [\underline{x}_i, \underline{x}_i^\mathcal{F}(j)) =: \mathcal{P}^\text{pos} \subseteq \mathcal{P}^\text{syn}
\end{align}
with some $\underline{x}_i^\mathcal{F}(j) > \underline{x}_i$ for $i\neq j$.

Note that this condition is robust to heterogeneity in the sense that one can preserve~\eqref{eq:exp_suff_time_cond} under a slight perturbation on the values $x_i''$, $\overline{x}_i^\text{mid}$, $\overline{x}_i$, and the function $h_i$.
Moreover, note that this robustness becomes stronger with respect to the monotonicity (Assumption~\ref{assum:mon_in_mean}), as when $\overline{H}_i(\cdot)$ is fixed, by decreasing $c_{\mathcal{S}\to\mathcal{F}}$ or $c_{\mathcal{F}\to\mathcal{S}}$, the speed of travel right after the spike, $|h_i(x)|$ at $x \in [\overline{x}_i^\text{mid}, \overline{x}_i]$, becomes faster relative to the speed of travel right before the spike, $|h_i(x)|$ at $x \in [\underline{x}_i, x_i'') \subset [\underline{x}_i, \underline{x}_i^\text{mid}]$.
Indeed, if individual neurons are identical, in particular, $h(\cdot) \equiv h_i(\cdot)$, $\underline{x} \equiv \underline{x}_i$, $\overline{x} \equiv \overline{x}_i$, $\underline{x}^\text{mid} \equiv \underline{x}_i^\text{mid}$, and $\overline{x}^\text{mid} \equiv \overline{x}_i^\text{mid}$ for all $i \in \mathcal{N}$, then the sufficient condition~\eqref{eq:exp_suff_time_cond} simply becomes
$$\tau(\overline{x}) - \tau(\underline{x}^\mathcal{F}) < \tau(\overline{x}^\text{mid}),$$
where $\underline{x}^\mathcal{F} := \min_{j \in \mathcal{N}} \max_{p'' \in \mathcal{B} \text{ s.t. } j(p'') = j} \min_{i \neq j} x_i''$.
This is equivalent to $\tau(\overline{x}) - \tau(\overline{x}^\text{mid}) < \tau(\underline{x}^\mathcal{F})$, or further to
$$\int_{\overline{x}}^{\overline{x}^\text{mid}} \frac{1}{h(x)} dx < \int_{\underline{x}^\mathcal{F}}^{\underline{x}} \frac{1}{h(x)} dx.$$
So, if the function $h(\cdot)$ is monotone in the mean:
$$\max_{x\in[\underline{x}, \underline{x}^\mathcal{F}]} |h(x)| \le \max_{x \in [\underline{x}, \underline{x}^\text{mid}]} |h(x)| \le c  \min_{x \in [\overline{x}^\text{mid},\overline{x}]} |h(x)|,$$
then we only need the monotonicity to be strong enough:
$$c < \frac{\underline{x}^\mathcal{F} - \underline{x}}{\overline{x} - \overline{x}^\text{mid}}.$$

Now, in the next subsection, we emphasize the important role of the existence of an open neighborhood around the fixed point in leading to rapid global convergence.

\subsection{Rapid global convergence}\label{subsec:global_convergence}

We first emphasize that under the sufficient condition given in the previous subsection, $\mathcal{P}^\text{syn}$ also corresponds to a network behavior of persistent synchronous spiking.
Now, this network behavior becomes globally stable if there is strong monotonicity: the speed of travel is sufficiently fast outside the neighborhood of the lower threshold.
This is intuitive as the neurons that have not spiked yet `wait' (by moving very slowly) for the neurons that have spiked, which significantly reduces the phase difference, eventually ending up synchronously firing.

\begin{thm1}[Stability inside $\mathcal{P}^\text{sp}$]\label{thm:global_convergence}
Given the integrate-and-fire model~\eqref{eq:i_n_f} with Properties~\ref{property:low_thres}--\ref{property:stab_spik}, in addition to Assumptions~\ref{assum:connec} and~\ref{assum:mon_in_mean} assume that there exists a positively invariant set $\mathcal{P}^\text{pos}$ of the form~\eqref{eq:def_P_pos} with $\underline{x}_i^\mathcal{F}(j) > \underline{x}_i$, $i \neq j$ (e.g., by assuming~\eqref{eq:exp_suff_time_cond} for all $j$), and
\begin{itemize}
\item (Sufficient monotonicity) there is $r_i \in (0, 1)$ so that 
$$\tau_i(\overline{x}_i) - \tau_i((1-r_i)\underline{x}_i + r_i\underline{x}_i^\mathcal{F}) < \tau_j(\overline{x}_j^\text{mid}), \quad \forall i \neq j,$$
where $\underline{x}_i^\mathcal{F} := \min_{j \neq i} \underline{x}_i^\mathcal{F}(j)$, hence the fixed point $p^* = {\rm col}(x_1^*, \dots, x_N^*)$ satisfies $x_i^* - \underline{x}_i <r_i(\underline{x}_i^\mathcal{F} - \underline{x}_i)$ for all $i \in \mathcal{N}$, and there exists $r_\eta \ge 0$ and $\eta^* \in [0, \min_{i \in \mathcal{N}} (1-r_i)[\underline{x}_i^\mathcal{F} - \underline{x}_i])$ so that
\begin{align*}
&(N-1)  D_{\mathcal{P}^\text{sp}} \\
&\quad\quad\quad < (1 -c_{\mathcal{S}\to\mathcal{F}} \overline{r} c_{\mathcal{F}\to\mathcal{S}})  d_{\mathcal{P}^\text{syn}} - r_\eta  (d_{\mathcal{P}^\text{sp}} - d_{\mathcal{P}^\text{syn}}),
\end{align*}
where $D_{\mathcal{P}^\text{sp}} :=  2  \max_{i \in \mathcal{N}}[\tau_i(\overline{x}_i) - \tau_i(\underline{x}_i^\mathcal{F} - \eta^*)]$, $d_{\mathcal{P}^\text{syn}} := \min_{i \in \mathcal{N}} [\tau_i(\underline{x}_i^\mathcal{F} - \eta^*) - \tau_i((1-r_i)\underline{x}_i + r_i\underline{x}_i^\mathcal{F})]$, and $d_{\mathcal{P}^\text{sp}} := \max_{i \in \mathcal{N}} [\tau_i(\overline{x}_i) - \tau_i(\underline{x}_i)]$.\footnote{Note that $D_{\mathcal{P}^\text{sp}} / d_{\mathcal{P}^\text{syn}}$ can be made arbitrarily small if the monotonicity is further extended so that there is a sufficient difference in the speed in the region $[(1-r_i)\underline{x}_i+r_i\underline{x}_i^\mathcal{F}, \underline{x}_i^\mathcal{F} - \eta^*)$ and $(\underline{x}_i^\mathcal{F} - \eta^*, \overline{x}_i]$.}
\end{itemize}
Then, for sufficiently small $\eta$: $\eta \in [0, \eta^*]$ that satisfies\footnote{Note that the left-hand side of the inequality can be made arbitrarily close to $1$ by decreasing $\eta$.}
\begin{align}\label{eq:eta}
\max_{i \in \mathcal{N}} \max_{x_i \in [\underline{x}_i, \overline{x}_i]}\max_{x_i^+ \in [\max\{x_i - \eta, \underline{x}_i\}, x_i]} \frac{h_i(x_i)}{h_i(x_i^+)}  \le  1+ r_\eta,
\end{align}
any trajectory that starts from $\mathcal{P}^\text{sp}$ enters the positively invariant set $\mathcal{P}^\text{syn}$ of persistent synchronous spiking within a couple of cycles (rapid convergence).
\QEDB
\end{thm1}

\begin{pf}
By~\eqref{eq:eta}, analogous to the argument in the proof of Theorem~\ref{thm:local_stab_i_n_f_fixed}, we can still conclude
$$\max_{i \in \mathcal{N}} \delta \tau_i^{++} - \min_{i \in \mathcal{N}} \delta \tau_i^{++} \le (1 + r_\eta)  \left[\max_{i \in \mathcal{N}} \delta\tau_i- \min_{i \in \mathcal{N}} \delta \tau_i\right]$$
at point $p = {\rm col}(x_1, \dots, x_N)$ in $\mathcal{P}^\text{sp}\setminus\mathcal{P}^\text{syn}$, for any infinitesimal perturbation ${\rm col}(\delta x_1, \dots, \delta x_N)$ that preserves $\mathcal{N}^\mathcal{F}(p)$ and an index $j$ such that $x_j = \underline{x}_j$.

This means that if there is a path that connects $p$ and $p'$ inside $\mathcal{P}^\text{sp}$ such that on the path, we preserve $\mathcal{N}^\mathcal{F}(p)$ and an index $j$ such that $x_j = \underline{x}_j$, then we can conclude
$$d(X^{++}(X^+(p)), X^{++}(X^+(p'))) \le (1 + r_\eta)  d(p, p'),$$
analogous to Theorem~\ref{thm:local_stab_i_n_f}.

Now, when this does not apply is when the path passes a point $p''$ where the index set of synchronously firing neurons changes.
However, we still know that
$$\lim_{p_+ \to p''+} \lim_{p_- \to p''-} d(X^{++}(X^+(p_+)), X^{++}(X^+(p_-)))$$
is bounded by a positive constant $\mathcal{D}_{\mathcal{P}^\text{sp}}$.
In particular,
\begin{align*}
&\lim_{p_+ \to p''+} \lim_{p_- \to p''-} d(X^{++}(X^+(p_+)), X^{++}(X^+(p_-))) \\
&\quad\quad= \lim_{p_+ \to p''+} \lim_{p_- \to p''-} \max_{i \in \mathcal{N}} [\tau_i(x_i^+(p_+)) - \tau_i(x_i^+(p_-))] \\
&\quad\quad\quad - \lim_{p_+ \to p''+} \lim_{p_- \to p''-} \min_{i \in \mathcal{N}} [\tau_i(x_i^+(p_+)) - \tau_i(x_i^+(p_-))] \\
&\quad\quad\le 2  \max_{i \in \mathcal{N}^\mathcal{F}(p_+) \triangle \mathcal{N}^\mathcal{F}(p_-)} [\tau_i(\overline{x}_i) - \tau_i(\underline{x}_i^\mathcal{F} - \eta)] \le D_{\mathcal{P}^\text{sp}},
\end{align*}
where $\mathcal{N}_+ \triangle \mathcal{N}_-$ denotes the symmetric difference between $\mathcal{N}_+$ and $\mathcal{N}_-$, i.e., $(\mathcal{N}_+\setminus\mathcal{N}_-) \cup (\mathcal{N}_- \setminus\mathcal{N}_+)$.

This finally implies that on the path that connects $p^*$ and $p' \in \mathcal{P}^\text{sp} \setminus \mathcal{P}^\text{syn}$, as in the proof of Theorem~\ref{thm:local_stab_i_n_f}, for any $\eta' > 0$, there exists $\hat{p} \in {\mathcal{P}^\text{syn}}$ such that $d(\hat{p}, p') \le d(p^*, p') - d(p^*, \hat{p}) + \eta'$ and $d(p^*, \hat{p}) \ge d_{\mathcal{P}^\text{syn}}$, which satisfies
\begin{align*}
&d(p^*, X^{++}(X^+(p'))) = d(X^{++}(X^+(p^*)), X^{++}(X^+(p'))) \\
&\le d(X^{++}(X^+(p^*)), X^{++}(X^+(\hat{p}))) \\
&\quad + d(X^{++}(X^+(\hat{p})), X^{++}(X^+(p'))) \\
&\le c_{\mathcal{S}\to\mathcal{F}}  \overline{r} c_{\mathcal{F}\to\mathcal{S}}  d(p^*, \hat{p}) + (N-1)  D_{\mathcal{P}^\text{sp}}  + (1+r_\eta) d(\hat{p}, p').
\end{align*}
This is because, by Property~\ref{property:convex}, for a monotonically increasing (decreasing) path, the points where the index set of synchronously firing neurons changes are isolated, hence there are at most $N-1$ of them.
By (Sufficient monotonicity), there exists a sufficiently small $c \in (0, 1)$ such that
\begin{align*}
(N-1) D_{\mathcal{P}^\text{sp}} &< (1 + r_\eta -c_{\mathcal{S}\to\mathcal{F}}  \overline{r} c_{\mathcal{F}\to\mathcal{S}})  d_{\mathcal{P}^\text{syn}} \\
&\quad - (c + r_\eta) d_{\mathcal{P}^\text{sp}}
\end{align*}
and this implies with the former argument that
\begin{align*}
d(p^*, X^{++}(X^+(p'))) &\le -(1 + r_\eta -c_{\mathcal{S}\to\mathcal{F}}  \overline{r} c_{\mathcal{F}\to\mathcal{S}})  d_{\mathcal{P}^\text{syn}} \\
&\quad + (N-1) D_{\mathcal{P}^\text{sp}} \\
&\quad + (1+r_\eta)  d(p^*, p') + (1+r_\eta) \eta'\\
&< (1 - c)  d(p^*, p') + (1+r_\eta)\eta'.
\end{align*}
As this holds for any $\eta' > 0$, we can conclude the proof.

\QEDBB
\end{pf}

A significant difference from previous studies is our focus on the spiking map from $\mathcal{P}^\text{syn}$ (local), contrary to the usual focus on the spiking map from $\mathcal{P}^\text{sp} \setminus \mathcal{P}^\text{syn}$ (global).
In particular, by departing from the assumptions of identical neurons and all-to-all coupling, we focused our attention on the behavior of persistent synchronous spiking itself, more precisely the existence of a hyperbolic fixed point.
It is the robustness of this attractor in the singular limit that allows us to conclude from singular perturbation analysis away from the singular limit.
As in previous studies~\citep{mauroy2011dichotomic}, the monotonicity (in the mean) is found essential.
In this subsection, we further remarked that global convergence property can be recovered by strong monotonicity.

The biophysical model illustrated in Section~\ref{sec:model} is standard, and thus, the integrate-and-fire model that we proposed in Section~\ref{sec:i_n_f} is well suited for the study of network behaviors in a neuronal network.
For instance, it provides robustness to the phenomenon of absorptions, motivated by the biophysical model.

\section{Main result: biophysical model}\label{sec:main_bio}

\subsection{Discussion}\label{subsec:discuss}

\begin{enumerate}
\item In the first part of Section~\ref{sec:main}, the assumption on the monotonicity in the mean (Assumption~\ref{assum:mon_in_mean}) has been shown to be sufficient for the local stability of the fixed point and the contraction inside $\mathcal{P}^\text{syn}$, which also gives uniqueness and the existence of a fixed point (given the positive invariance of $\mathcal{P}^\text{syn}$).
For the biophysical model of Section~\ref{sec:model}, $h_i(\cdot)$ ($\overline{H}_i(\cdot)$) in Assumption~\ref{assum:mon_in_mean} is determined by the slow (fast) behavior on the lower (upper) branch.
Thus, as first observed in~\citep{somers1993rapid}, the key property is the ``rate of compression,'' which measures relative velocities of the slow variable just before and just after an ultrafast-fast jump of the neuron, the parameters $c_{\mathcal{S}\to\mathcal{F}}$ and $c_{\mathcal{F}\to\mathcal{S}}$ in Assumption~\ref{assum:mon_in_mean}.
In particular, Assumption~\ref{assum:mon_in_mean} is the exact counterpart of the stability condition given in~\citep{somers1993rapid}, especially the weaker one mentioned in its Remark 2.5, while in~\citep{somers1993rapid} systems with a two-time scale (slow=fast-ultrafast) are considered instead.
In the proof of Theorem~\ref{thm:local_stab_i_n_f_fixed}, we have explained how such a compression condition arises in an analytical treatment of simple linearization, even for heterogeneous networks. 
This property is shared by biophysical conductance-based models of neuronal systems~\citep{somers1993rapid}.

\item We emphasize that, for the biophysical model of Section~\ref{sec:model}, Assumption~\ref{assum:mon_in_mean} is completely determined by the individual behavior of neurons: the shape of the individual nullcline, the time constant, and the robustness margin.
In particular, a sufficient condition for Assumption~\ref{assum:mon_in_mean} is that their slow trajectory has a ``scalloped'' shape~\citep{somers1993rapid}: that there exists a sufficiently small $\lambda < 0$ such that 
$$h_i'(x) \le \lambda < 0, \quad \forall x \in (\underline{x}_i(0), \overline{x}_i(M_i)), \quad \forall i \in \mathcal{N}.$$
This means that the properties of the individual neurons determine the rate of convergence to synchrony, independent of the coupling strength and the network topology (however, they may affect the size or the positive invariance of $\mathcal{P}^\text{syn}$, see the next item).

\item In Section~\ref{subsubsec:exist_P_per}, an explicit sufficient condition~\eqref{eq:exp_suff_time_cond} for the existence of a fixed point has been found based on the approximation~\eqref{eq:approx_P_syn}.
According to Section~\ref{subsec:sync_jump}, for the biophysical model illustrated in Section~\ref{sec:model}, especially by Property~\ref{propertyp:FTM}, $\mathcal{P}^\text{syn}$ can be exactly characterized with the collection $\mathcal{B}$ of points $p^{\mathcal{B}_\mathcal{T}}$ indexed by a spanning tree $\mathcal{T}$ of $\mathcal{G}$.
Also, by Property~\ref{propertyp:spike_map}, $\overline{x}_i$ can be relaxed as $\overline{x}_i(g_id_i)$, and thus, the explicit sufficient condition~\eqref{eq:exp_suff_time_cond} becomes
\begin{align}\label{eq:suff_cond_bio_model}
\begin{split}
&\exists \text{ spanning tree }\mathcal{T} \text{ of } \mathcal{G} \text{ having a root } j \text{ s.t. } \\
&\tau_i(\overline{x}_i(g_id_i)) - \tau_i(\underline{x}_i(g_i\overline{d}_i^\mathcal{T})) < \tau_j(\overline{x}_j(0)), \quad \forall i  \neq  j,
\end{split}
\end{align}
for all $ j\in \mathcal{N}$ such that $\tau_j(\overline{x}_j(0)) \le \min_{i \in \mathcal{N}} \tau_i(\overline{x}_i(g_id_i))$.
In particular, this is equivalent to whether for each $j \in \mathcal{N}$ the resulting set $\mathcal{N}^\mathcal{T}$ of the following algorithm is equal to the entire set $\mathcal{N}$.
\begin{enumerate}
\item Let $l = 1$ and $\mathcal{N}_0^\mathcal{T} = \{j\}$.
\item Let $\mathcal{N}_l^\mathcal{T} = \{i : \tau_i(\overline{x}_i(g_id_i)) - \tau_j(\overline{x}_j(0)) < \tau_i(\underline{x}_i(g_i\overline{d}_i^\mathcal{T}))\} \setminus \overline{\mathcal{N}}_{l-1}^\mathcal{T}$, where $\overline{d}_i^\mathcal{T} := \sum_{k \in \mathcal{N}_i\cap \overline{\mathcal{N}}_{l-1}^\mathcal{T}} \alpha_{ik}$.
\item If $\mathcal{N}_l^\mathcal{T} \neq \emptyset$, then let $l = l+1$ and repeat (b).
Otherwise, let $l^* = l-1$ and define $\mathcal{N}^\mathcal{T} := \overline{\mathcal{N}}_{l^*}^\mathcal{T}$.
\end{enumerate}
As noted in Section~\ref{subsubsec:exist_P_per}, this is easier to satisfy if the rate of compression is small.

Note that this can even be satisfied for arbitrarily weak coupling strength (when $g_i \to 0$), under the assumption that there exists $\overline{T} > 0$ such that
$$\overline{T} = \tau_i(\overline{x}_i(0)) - \tau_i(\underline{x}_i(0)) = \tau_i(\overline{x}_i(0)), \quad \forall i \in \mathcal{N},$$
meaning that all neurons should have an identical period for their decoupled limit cycle.
This is also necessary.
Under this additional condition, our sufficient condition~\eqref{eq:suff_cond_bio_model} can be guaranteed for sufficiently small $g_\text{coup}$ (where $g_i = g_\text{coup}\overline{g}_i$) if and only if
\begin{align}\label{eq:suff_cond_arb_small}
\begin{split}
&\exists \text{ spanning tree }\mathcal{T} \text{ of } \mathcal{G} \text{ having a root } j \text{ s.t. } \\
&\tau_i'(\overline{x}_i(0)) \overline{x}_i'(0) \overline{g}_id_i < \tau_i'(\underline{x}_i(0)) \underline{x}_i'(0) \overline{g}_i\overline{d}_i^\mathcal{T}
\end{split}
\end{align}
for all $i \neq j$ and $j \in \mathcal{N}$.
Again this is easier to satisfy if the rate of compression is small.
In fact, this is always satisfied for biophysical conductance-based models if the graph is sufficiently balanced, i.e., $d_i = \min_{j \neq i} \min_{\mathcal{T}, \text{root} =j} \overline{d}_i^\mathcal{T} =: \underline{d}_i$, because, by monotonicity in the mean, we have
$$\tau_i'(\overline{x}_i(0)) = -\frac{1}{h_i(\overline{x}_i(0))} < -\frac{1}{h_i(\underline{x}_i(0))} = \tau_i'(\underline{x}_i(0))$$
and biophysical conductance-based models satisfy\footnote{The equalities can be derived from the identities $(d/dm)f_i^m(\underline{x}_i(m), \underline{v}_i(m)) = (d/dm)f_i^m(\overline{x}_i(m), \overline{v}_i(m)) = 0$ and $(\partial f_i^m/\partial v)(\underline{x}_i(m), \underline{v}_i(m)) = (\partial f_i^m /\partial v)(\overline{x}_i(m), \overline{v}_i(m)) = 0$.}
\begin{align*}
\overline{x}_i'(0) &= \frac{\overline{E}_i - \overline{v}_i(0)}{-\frac{\partial f_i}{\partial x}(\overline{x}_i(0), \overline{v}_i(0))} \\
&\,\,\,\,\quad\quad\quad\quad\quad\quad \le \frac{\overline{E}_i - \underline{v}_i(0)}{-\frac{\partial f_i}{\partial x}(\underline{x}_i(0), \underline{v}_i(0))} = \underline{x}_i'(0),
\end{align*}
e.g., for conductance-based models as in Sections~\ref{sec:me} and~\ref{sec:sim}, we have $(-\partial f_i/\partial x)(x, v) = g_{i,K}(0.7+v) > 0$.
A similar argument shows that if the conditions on monotonicity and conductance-based models are extended for $g \in (0, {g}_\text{coup})$, then~\eqref{eq:suff_cond_bio_model} is satisfied.

An example of a sufficiently balanced graph with unitary weights, i.e., $\alpha_{ij} = 1$ for all $(j, i) \in \mathcal{E}$, is the directed ring network, where $\mathcal{N}_i = \{i+1\}$ for all $i \neq N$ and $\mathcal{N}_N = \{1\}$ ($d_i = \underline{d}_i = 1$).
This conclusion supports the numerical observation at the end of Section 3.1 of~\citep{somers1993rapid}.
Indeed, sparsity in connection reduces the difference between ${d}_i$ and $\underline{d}_i$, which further favors our sufficient condition~\eqref{eq:suff_cond_arb_small}.

\item We emphasize again that the strong contraction provided by the nodal excitability (given by Assumptions~\ref{assum:spike}--\ref{assum:robust}) and the \emph{weak} localized synaptic coupling (defined by Assumptions~\ref{assum:weak} and~\ref{assum:local}) is what gives \emph{rapid} convergence by constructing an open set $\mathcal{P}^\text{syn}$ of persistent synchronous spiking around the fixed point $p^*$, and thus, maintains the convergence rate $c_{\mathcal{S}\to\mathcal{F}}\overline{r} c_{\mathcal{F}\to\mathcal{S}}$ even when the weak coupling strength $g_i$ is further reduced.
These assumptions are \emph{robust} to heterogeneity. 

\item Theorem~\ref{thm:global_convergence} in Section~\ref{subsec:global_convergence}, underlines the important role of the domain of attraction $\mathcal{P}^\text{syn}$ having a volume in leading to rapid global convergence.
In particular, the global convergence property is recovered by sufficient compression: the speed of slow travel is sufficiently fast outside the neighborhood of the left knee.
This is consistent with the numerical observation in item 1 of Section 3.3 of~\citep{somers1993rapid}.

\item Now, the conclusion of Theorem~\ref{thm:global_convergence} results in an almost global convergence towards the fixed point $p^*$ with the help of the first step of the convergence illustrated in the Introduction: rapid convergence (finite-time convergence for the $\epsilon = 0$ limit) to $\mathcal{P}^\text{sp}$, which is obtained by preserving the biophysical nature of each individual neuron by weak coupling.
See Appendix~\ref{app:almost_global} for the complete argument.
This leads to our main theorem illustrated in the next subsection.
\end{enumerate}

\subsection{Main theorem}

\begin{thm1}[Rapid and robust synchronization]\label{cor:eps}
In addition to Assumptions~\ref{assum:spike}--\ref{assum:robust},~\ref{assum:weak}--\ref{assum:local},~\ref{assum:connec}, and~\ref{assum:mon_in_mean}, assume~\eqref{eq:suff_cond_bio_model} for all $j \in \mathcal{N}$, and 
\begin{itemize}
\item (Sufficient compression) there is $r_i \in (0, 1)$ so that
$$\tau_i(\overline{x}_i(g_id_i)) - \tau_i(\underline{x}_i(r_i  g_i\underline{d}_i)) < \tau_j(\overline{x}_j(0)), \quad \forall i \neq j,$$
hence the fixed point $p^* = {\rm col}(x_1^*, \dots, x_N^*)$ satisfies $x_i^* \in [\underline{x}_i(0), \underline{x}_i(r_i  g_i\underline{d}_i))$ for all $i \in \mathcal{N}$, and
$$(N-1) D_{\mathcal{P}^\text{sp}} < (1 - c_{\mathcal{S} \to \mathcal{F}}  \overline{r}  c_{\mathcal{F}\to\mathcal{S}})  d_{\mathcal{P}^\text{syn}},$$
where $D_{\mathcal{P}^\text{sp}} := 2 \max_{i \in \mathcal{N}} [\tau_i(\overline{x}_i(g_id_i)) - \tau_i(\underline{x}_i(g_i\underline{d}_i))]$ and $d_{\mathcal{P}^\text{syn}} := \min_{i \in \mathcal{N}} [\tau_i(\underline{x}_i(g_i\underline{d}_i)) - \tau_i(\underline{x}_i(r_i  g_i\underline{d}_i))]$.
\end{itemize}
Then, there exists $\epsilon^* > 0$ such that for each $\epsilon \in (0, \epsilon^*)$, the network~\eqref{eq:ind_neu} with~\eqref{eq:syn} has an almost semi-globally stable limit cycle, that corresponds to a persistent synchronous spiking network behavior, which has rapid convergence.
In particular,
\begin{enumerate}
\item by preserving the biophysical nature of individual neurons by weak coupling (Assumption~\ref{assum:weak}), the network rapidly converges to a neighborhood of $\mathcal{S}$, i.e., each neuron rapidly converges to a neighbor of its limit cycle attractor. 
Then, after less than a cycle, it converges into a neighborhood of $\mathcal{P}^\text{sp}$.
\item After a couple of cycles, this results in a neighborhood of $\mathcal{P}^\text{syn}$, a positively invariant set that corresponds to a persistent synchronous spiking network behavior, as presynaptic neurons only affect the postsynaptic neuron when synchronously spiking (the third step of the convergence).
\item Finally, these network behaviors of persistent synchronous spiking converge to the unique limit cycle with a convergence rate that is independent of the coupling strength and the network topology (the second step of the convergence).\QEDB
\end{enumerate}
\end{thm1}

Theorem~\ref{cor:eps} follows from Theorems~\ref{thm:local_stab_i_n_f}--\ref{thm:global_convergence} and from the arguments in Sections~\ref{subsec:sync_jump},~\ref{subsec:spik_map},~\ref{subsec:discuss}, and Appendices~\ref{app:just}--\ref{app:almost_global}.
This emphasizes that the network under study achieves \emph{rapid} convergence to a persistent \emph{synchronous} spiking network behavior that is \emph{robust} to heterogeneity via \emph{weak synaptic} coupling.

\begin{rem1}
Theorem~\ref{cor:eps} proves the conjecture of Somers and Kopell in Section 3.4 of~\citep{somers1993rapid}.
\QEDB
\end{rem1}

\section{Simulation}\label{sec:sim}

To illustrate Assumptions~\ref{assum:spike}--\ref{assum:robust},~\ref{assum:weak}--\ref{assum:local},~\ref{assum:connec},~\ref{assum:mon_in_mean}, and especially the robustness of the existence of $p^*$ with respect to heterogeneity, we simulate a network consisting of the type of systems given as
\begin{align*}
\tau_{\epsilon}(v_i) \dot{x}_i &= -x_i + x_{\infty}(v_i) \\ 
\epsilon\dot{v}_i &= g_{i,\text{L}}(-0.4 - v_i) + g_{i,\text{Ca}} m_\infty(v_i)(1 - v_i ) \\
&\quad+ g_{i,\text{K}}x_i(-0.7 - v_i)  +  0.4\\ 
&\quad + g_\text{coup} (1 - v_i)\sum_{j \in \mathcal{N}_i} S(v_j)
\end{align*}
where $\epsilon = 0.0004$ and $m_\infty(\cdot)$ is given in Section~\ref{sec:me}, which resembles the standard Morris-Lecar excitable model in Section~\ref{sec:me}.
Here, heterogeneous parameters are the maximal conductances $g_{i,\text{L}}$, $g_{i,\text{Ca}}$, and $g_{i,\text{K}}$, as in the neuronal network.
We take the parameters random with uniform distribution from the intervals $[0.3, 0.75]$, $[1.75, 2.25]$, and $[2.75, 3.25]$, respectively.
If $x_{\infty}(\cdot)$ is a sigmoidal function ranging in $[0, 1]$, then we see that $\mathcal{D}_i = [0, 1] \times [-0.7, 1]$ is positively invariant for any $m \ge 0$, as
\begin{align*}
f_i^m(x, v) &= g_{i,\text{L}}(-0.4 -v) + g_{i, \text{Ca}}m_\infty(v)(1 - v) \\
&\quad + g_{i,\text{K}}x(-0.7-v)  + 0.4 + (1 - v)m\\
&< -1.4 g_{i,\text{L}} + 0.4 \le -0.02 < 0
\end{align*}
for all $x \ge 0$ and $v > 1$, while $f_i^m(x, v) > 0.4 > 0$ for all $x \ge 0$ and $v < -0.7$.

The nullcline $f_i^m(x, v) = 0$ has $N$-shape if 
\begin{align*}
&[0.3g_{i, \text{L}} + 1.7m + 0.4]/g_{i, \text{Ca}} \\
&\,\,\,\,\quad< 0.71*0.99*m_\infty'(0.01) - 1.7 m_\infty(0.01) = 1.4260, \\
&\,\,\,\,\quad< 0.7*1*m_\infty'(0) - 1.7 m_\infty(0) = 1.4833.
\end{align*}
This can be deduced by calculating the existence condition of two zero roots for the partial derivative of the nullcline $f_i^m(x, v) = 0$ with respect to $v$.
In fact, this is the condition that guarantees $(\partial x_i^m/\partial v)(0.01), (\partial x_i^m/\partial v)(0) > 0$, where $x_i^m(v)$ is given in Lemma~\ref{lem:move_right}.
Thus, we take our robustness margin as
\begin{align*}
M_i := 0.36 <  1.1003 &= \frac{1.4260* 1.75- 0.3 * 0.75 - 0.4}{1.7} \\
&\le \frac{1.4260 * g_{i,\text{Ca}} - 0.3 * g_{i,\text{L}} - 0.4}{1.7}.
\end{align*} 
Note that $(\partial x_i^m/\partial v)(0.01), (\partial x_i^m/\partial v)(0) > 0$ also tells us that the horizontal lines $v = 0.01$ and $v = 0$ are always located between the left knee and the right knee.
Then, Assumptions~\ref{assum:spike} and~\ref{assum:robust} are satisfied if $x_{\infty}(\cdot)$ is $0$ in the lower branch and $1$ in the upper branch.
This is because, we have
\begin{align*}
\underline{x}_i(M_i) &< x_i^{M_i}(v' := -0.1460) \\
&\le x_i^0(v'' := 0.0930) < \overline{x}_i(0),
\end{align*}
where the second inequality holds if and only if
\begin{align*}
M_i(1-v')(0.7+v'') &\le -(0.3g_{i, \text{L}}+0.4)(v'' - v') \\
&\quad - g_{i, \text{Ca}}m_\infty(v')(1-v')(0.7+v'') \\
&\quad + g_{i, \text{Ca}} m_\infty(v'')(1-v'')(0.7+v').
\end{align*}
Moreover, $0 < \underline{x}_i(m) < \overline{x}_i(m) < 1$ because
\begin{align*}
f_i^m(0, v) &\ge \min\{ 0.3 (-0.4-v), 0.75(-0.4-v)\} \\
&\quad + 1.75m_\infty(v)(1-v) + 0.4 > 0, \\
&\,\,\,\,\,\quad\quad\quad\quad\quad\quad\quad\quad\quad\quad\quad\quad\quad\quad v \in [-0.7, 0], \\
f_i^m(1, v) &\le 0.3(-0.4-v) + (2.25m_\infty(v)+0.36)(1-v) \\
&\quad + 2.75(-0.7-v) + 0.4 < 0, \,\,\,\quad  v \in [0.01, 1].
\end{align*}
To satisfy the rest of the conditions in Assumptions~\ref{assum:bio},~\ref{assum:robust}, and~\ref{assum:local}, we model $\tau_{\epsilon}(\cdot)$, $x_{\infty}(\cdot)$, and $S(\cdot)$ as 
\begin{align*}
x_{\infty}(v) &= S(v) = \begin{cases}1, &\mbox { if } v \ge 0, \\ 0, &\mbox{ if } v < 0,\end{cases}\\
\tau_{\epsilon}(v) &= \begin{cases}\epsilon^{q}, &\mbox{ if } v \ge E^\text{th}, \\ \epsilon^{q} + \underline{\tau}, &\mbox{ if } v \in [-0.258, E^\text{th}), \\
\epsilon^q + \overline{\tau}, &\mbox{ if } v < -0.258,\end{cases}
\end{align*}
with $q = 1/2$, $E^\text{th} = 0.01$, and $\overline{\tau} = 5 \gg \epsilon^q$. 
Note that the horizontal line $v = -0.258$ always has an intersection with the lower branch (we always have $(\partial x_i^m/\partial v)(-0.258) < 0$), hence the rate of compression can be approximated as
$$\frac{\sup_{x \in [\underline{x}_i, x_i^m(-0.258))} |h_i(x)|}{\inf_{x \in (x_i^m(-0.258),\overline{x}_i]}|h_i(x)|} = \frac{\overline{\tau}}{\underline{\tau}} = \frac{5}{\underline{\tau}}.$$

We utilize a graph satisfying Assumption~\ref{assum:connec}, where we selected $36$ neighbors for each $i$ as $i+1, \dots, i+36 \mod N$, where the total number of neurons is $N = 100$.
By selecting adjacency elements in $\{0, 1\}$, we get $d_i = 36$ for all $i \in \mathcal{N}$.
Hence, we take our weak coupling strength as $g_\text{coup} = 0.01$ to satisfy Assumption~\ref{assum:weak} with the robustness margin $M_i = 0.36$.
Then, we can see from the following simulation results that as the rate of compression $c$ gets smaller: as $\underline{\tau} > 0$ gets larger, the conditions for the existence of $p^*$ and for the global convergence are further satisfied for the varying parameters $g_{i,\text{L}}$, $g_{i,\text{Ca}}$, and $g_{i,\text{K}}$ in a broad range.

\begin{figure}[h]
\begin{center}
\includegraphics[width=\columnwidth]{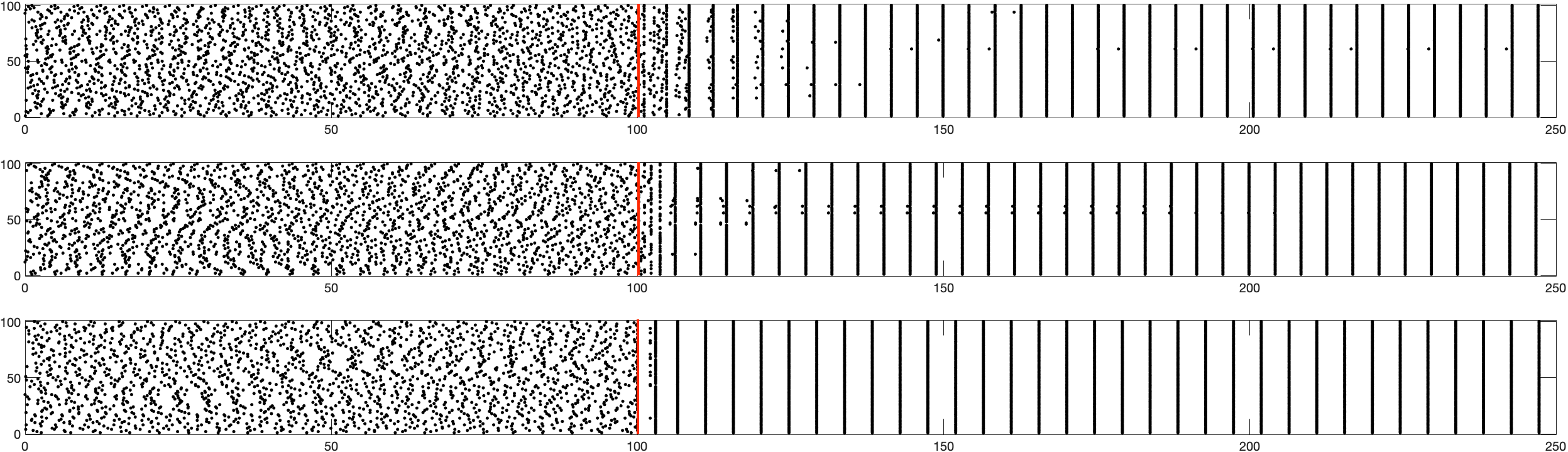}
\caption{A simulation result of $100$ neurons modeled in the manner explained in Section~\ref{sec:sim} is shown as a raster plot of spike times (i.e., times of the extreme peak for each cycle) of individual neurons, where the initial condition is also taken randomly with uniform distribution from $\mathcal{D}_i$.
The top, middle, and bottom figures correspond to $\underline{\tau} = 0.5$, $5$, and $30$, respectively.
For illustration, decoupled heterogeneous neurons are shown until $t = 100$ and the synaptic coupling becomes active at the red line.}
\label{fig:4}
\end{center}
\end{figure}

\section{Conclusion}\label{sec:conc}
We have illustrated how weak localized synaptic coupling combined with spiking oscillators can result in rapid spike synchronization in heterogeneous networks.
Traditionally, the analysis of synchronization in the vicinity of a limit cycle is achieved via a reduction to phase models (neuron-wise reductions).
This reduction however fails in the singular limit of slow-fast attractors. 
We propose instead that direct linearization of a network-wise reduction of a state-space conductance-based model provides an elucidating connection to the classical linear theory of network synchronization by diffusive coupling.
We showed the essential role of the ``rate of compression'' (monotonicity) for stability and robustness, which connects our analysis to the theory of synchronization in integrate-and-fire networks~\citep{mirollo1990synchronization,mauroy2008clustering,mauroy2011dichotomic}.
The mechanism analyzed in the present paper is of general nature in excitable models.
It would be of interest to extend the analysis to other types of synchronization, for instance, synchronization mechanisms in heterogeneous ensembles of bursting neurons.

\bibliographystyle{pre-automatica}
\bibliography{Reference}

\appendix

\section{A network-wise singular perturbation analysis}\label{app:just}

We first introduce a method to approximate singularly perturbed systems, in which the fast system is not uniformly stable, as in our network~\eqref{eq:ind_neu} with~\eqref{eq:syn}.
The method is to first normalize the entire dynamics and then consider the $\epsilon = 0$ limit.
In particular, for our network~\eqref{eq:ind_neu} with~\eqref{eq:syn}, we get the parametrized normalized network as
\begin{align}\label{eq:norm_net}
\begin{split}
\dot{\hat{x}}_i &= \frac{h_{i, \epsilon}(\hat{x}_i, \hat{v}_i)}{n_\epsilon(\hat{x}_1, \hat{v}_1, \dots, \hat{x}_N, \hat{v}_N)} \\
\dot{\hat{v}}_i &= \frac{\hat{f}_i(\hat{x}_i, \hat{v}_i, \{\hat{v}_j\}_{\mathcal{N}_i})/\epsilon}{n_\epsilon(\hat{x}_1, \hat{v}_1, \dots, \hat{x}_N, \hat{v}_N)}, \quad \forall i \in \mathcal{N}
\end{split}
\end{align}
where
\begin{align*}
h_{i, \epsilon}&:= \frac{-\hat{x}_i + x_{i, \infty}(\hat{v}_i)}{\tau_{i, \epsilon}(\hat{v}_i)}, \\
\hat{f}_i&:= f_i(\hat{x}_i, \hat{v}_i) + g_i\left(\overline{E}_i - \hat{v}_i\right)\sum_{j \in \mathcal{N}_i}\alpha_{ij}S_{ij}(\hat{v}_j), \\
n_\epsilon &:= \sqrt{\sum_{i=1}^N \left(h_{i, \epsilon}(\hat{x}_i, \hat{v}_i)^2 + (\hat{f}_i(\hat{x}_i, \hat{v}_i, \{\hat{v}_j\}_{\mathcal{N}_i})/\epsilon)^2\right)}.
\end{align*}

Now, by considering the $\epsilon = 0$ limit, we obtain a discontinuous dynamical system given as
\begin{align}\label{eq:dds_S}
\begin{split}
\dot{\hat{x}}_i = \frac{h_i(\hat{x}_i)}{n_\mathcal{S}(\hat{x}_1, \dots, \hat{x}_N)} \,\,\text{ and }\,\, \dot{\hat{v}}_i &= 0, \quad\forall i \in \mathcal{N}, \\
n_\mathcal{S}(\hat{x}_1, \dots, \hat{x}_N) &:= \sqrt{\sum_{i=1}^N h_i(\hat{x}_i)^2},
\end{split}
\end{align}
when ${\rm col}(\hat{x}_1, \hat{v}_1, \dots, \hat{x}_N, \hat{v}_N) \in \mathcal{S}$,
\begin{align}\label{eq:dds_UF}
\begin{split}
\dot{\hat{x}}_i = 0 \text{ and } \dot{\hat{v}}_i &= \frac{\hat{f}_i(\hat{x}_i, \hat{v}_i, \{\hat{v}_j\}_{\mathcal{N}_i})}{\hat{n}_\mathcal{UF}(\hat{x}_1, \hat{v}_1, \dots, \hat{x}_N, \hat{v}_N)}, \quad \forall i \in \mathcal{N}, \\
\hat{n}_\mathcal{UF} &:= \sqrt{\sum_{i=1}^N \hat{f}_i(\hat{x}_i, \hat{v}_i, \{\hat{v}_j\}_{\mathcal{N}_i})^2},
\end{split}
\end{align}
when ${\rm col}(\hat{x}_1, \hat{v}_1, \dots, \hat{x}_N, \hat{v}_N) \in \mathcal{UF}$, and 
\begin{align}\label{eq:dds_F}
\begin{split}
\dot{\hat{x}}_i = \frac{H_i^{m_i}(\hat{x}_i)}{n_\mathcal{F}(\hat{x}_1, \dots,  \hat{x}_N)} \text{ and }\dot{\hat{v}}_i &= 0, \quad \forall i \in \mathcal{N}^\mathcal{F}, \\
\dot{\hat{x}}_i = 0 \text{ and } \dot{\hat{v}}_i &= 0, \quad \forall i \notin \mathcal{N}^\mathcal{F}, \\
n_\mathcal{F}(\hat{x}_1, \dots, \hat{x}_N) &:= \sqrt{\sum_{i \in \mathcal{N}^\mathcal{F}} H_i^{m_i}(\hat{x}_i)^2},
\end{split}
\end{align}
when ${\rm col}(\hat{x}_1, \hat{v}_1, \dots, \hat{x}_N, \hat{v}_N) \in \mathcal{F}$.
Then, the Filippov solution of this network can be obtained by the corresponding differential inclusion~\citep{filippov2013differential}, which results in the sliding trajectory on $\mathcal{S}$ following the dynamics
\begin{align}\label{eq:slid_S}
\begin{split}
\dot{\hat{x}}_i &= \frac{h_i(\hat{x}_i)}{\hat{n}_\mathcal{S}(\hat{x}_1, \hat{v}_1, \dots, \hat{x}_N, \hat{v}_N)}, \\
\dot{\hat{v}}_i &= -\frac{(\partial f_i/\partial x)(\hat{x}_i, \hat{v}_i)}{(\partial f_i/\partial v)(\hat{x}_i, \hat{v}_i)}\frac{h_i(\hat{x}_i)}{\hat{n}_\mathcal{S}(\hat{x}_1, \hat{v}_1, \dots, \hat{x}_N, \hat{v}_N)} \\
&=: R_i(\hat{x}_i, \hat{v}_i) \frac{h_i(\hat{x}_i)}{\hat{n}_\mathcal{S}(\hat{x}_1, \hat{v}_1, \dots, \hat{x}_N, \hat{v}_N)}, \,\,\,\quad\quad\quad \forall i \in \mathcal{N}, \\
\hat{n}_\mathcal{S} &:= \sqrt{\sum_{i=1}^N \left(1 + R_i(\hat{x}_i, \hat{v}_i)^2\right)h_i(\hat{x}_i)^2}.
\end{split}
\end{align}
Note that at jump points, i.e., points on $\mathcal{P}^\text{sp}$ introduced in~\eqref{eq:large_Poincare}, if neuron~$i$ approaches the left knee, i.e, $\hat{x}_i \to \underline{x}_i(0)$, then we have $R_i(\hat{x}_i, \hat{v}_i) \to -\infty$, hence
$$R_i(\hat{x}_i, \hat{v}_i)\frac{h_i(\hat{x}_i)}{\hat{n}_\mathcal{S}(\hat{x}_1, \hat{v}_1, \dots, \hat{x}_N, \hat{v}_N)} = \dot{\hat{v}}_i \to 1$$
while $\dot{\hat{x}}_j, \dot{\hat{v}}_j \to 0$ for $j \neq i$ and $\dot{\hat{x}}_i \to 0$.
This implies that we smoothly transit to the region $\mathcal{UF}$, which follows~\eqref{eq:dds_UF}.
Similarly, the Filippov solution results in the sliding trajectory on $\mathcal{F}$ following the dynamics
\begin{align}\label{eq:slid_F}
\begin{split}
\dot{\hat{x}}_i &= \frac{H_i^{m_i}(\hat{x}_i)}{\hat{n}_\mathcal{F}(\hat{x}_1, \hat{v}_1, \dots, \hat{x}_N, \hat{v}_N)}, \\
\dot{\hat{v}}_i &= -\frac{(\partial f_i^{m_i}/\partial x)(\hat{x}_i, \hat{v}_i)}{(\partial f_i^{m_i}/\partial v)(\hat{x}_i, \hat{v}_i)}\frac{H_i^{m_i}(\hat{x}_i)}{\hat{n}_\mathcal{F}(\hat{x}_1, \hat{v}_1, \dots, \hat{x}_N, \hat{v}_N)} \\
&=: R_i^{m_i}(\hat{x}_i, \hat{v}_i) \frac{H_i^{m_i}(\hat{x}_i)}{\hat{n}_\mathcal{F}(\hat{x}_1, \hat{v}_1, \dots, \hat{x}_N, \hat{v}_N)}, \,\,\quad\quad i \in \mathcal{N}^\mathcal{F}, \\
\dot{\hat{x}}_i &= 0 \text{ and } \dot{\hat{v}}_i = 0, \,\,\,\,\,\quad\quad\quad\quad\quad\quad\quad\quad\quad\quad\quad i \notin \mathcal{N}^\mathcal{F}, \\
\hat{n}_\mathcal{F} &:= \sqrt{\sum_{i \in \mathcal{N}^\mathcal{F}} (1 + R_i^{m_i}(\hat{x}_i, \hat{v}_i)^2)H_i^{m_i}(\hat{x}_i)^2}.
\end{split}
\end{align}
If neuron $i \in \mathcal{N}^\mathcal{F}$ approaches the right knee, i.e., $\hat{x}_i \to \overline{x}_i(m_i)$, then we have $R_i^{m_i}(\hat{x}_i, \hat{v}_i) \to -\infty $, hence 
$$R_i^{m_i}(\hat{x}_i, \hat{v}_i)\frac{H_i^{m_i}(\hat{x}_i)}{\hat{n}_\mathcal{F}(\hat{x}_1, \hat{v}_1, \dots, \hat{x}_N, \hat{v}_N)} = \dot{\hat{v}}_i \to -1.$$
This again implies that we smoothly transit to $\mathcal{UF}$.

Notice that the obtained solution trajectory is a bit different from that of the network introduced in Sections~\ref{subsec:fs_net},~\ref{subsec:sync_jump}, and~\ref{subsec:spik_map}.
In particular,~\eqref{eq:net_fast} in Section~\ref{subsec:sync_jump} requires an infinite time of travel for a jump.
However, what we only care about for this dynamics is the shape of the trajectory; the resulting point after a jump, and its effect on infinitesimal perturbation.
In this respect, we could identify two networks, the one introduced in Sections~\ref{subsec:fs_net},~\ref{subsec:sync_jump}, and~\ref{subsec:spik_map} and the one introduced in this subsection.
This is because, the point before and after a jump is on the $N$-dimensional manifolds $\mathcal{S}$ and $\mathcal{F}$, which is completely determined by $x_i$'s.
Note that $x_i$'s and $\delta x_i$'s do not change during this period of travel.

On the other hand,~\eqref{eq:net_slow} (or~\eqref{eq:net_nor}) is also different from~\eqref{eq:slid_S} (or~\eqref{eq:slid_F}), since~\eqref{eq:net_slow} (or~\eqref{eq:net_nor}) becomes infinitely faster than~\eqref{eq:slid_S} (or~\eqref{eq:slid_F}) as it approaches the knee (for some $x_i$).
However, again we are only interested in its trajectory; the resulting point after slow (fast) travel, and its effect on infinitesimal perturbation.
In this respect, we could identify two networks, since we preserve the identity
\begin{align*}
&\int_{\hat{x}_i^-}^{\hat{x}_i^+} \frac{1}{h_i(x)}dx \\
&= \int_{\tau^-}^{\tau^+} \frac{1}{\hat{n}_\mathcal{S}(\hat{x}_1(\tau), \hat{v}_1(\tau), \dots, \hat{x}_N(\tau), \hat{v}_N(\tau))} d\tau \\
&= \int_{\hat{x}_j^-}^{\hat{x}_j^+} \frac{1}{h_j(x)} dx
\end{align*}
and similarly, the identity
\begin{align*}
\int_{\hat{x}_i^-}^{\hat{x}_i^+} \frac{1}{H_i^{m_i}(x)}dx &= \int_{\hat{x}_j^-}^{\hat{x}_j^+} \frac{1}{H_j^{m_j}(x)} dx,
\end{align*}
which preserves both the argument in Sections~\ref{subsec:spik_map} and~\ref{subsec:local} and the resulting point.

In particular, analysis performed in Sections~\ref{subsec:fs_net}--\ref{subsec:local} all applies to the network defined here.

Now, therefore, from Theorems~\ref{thm:local_stab_i_n_f_fixed}--\ref{thm:global_convergence}, we can conclude that the discontinuous dynamical system obtained in this subsection, also has a unique almost globally stable limit cycle, which is contained in a positively invariant set representing the behavior of persistent synchronous spiking, which has the convergence rate independent of the weak coupling strength and the network topology.

For sufficiently small $\epsilon$, then we can find a slightly smaller (but extended in the direction of ultrafast convergence) positively invariant set containing the unique stable limit cycle, which has rapid convergence.
Now, the semi-global convergence follows from the continuous dependence for any trajectory that has a finite length~\citep{filippov2013differential}.
The exception is made around the measure zero set $\cup_{\mathcal{N}' \subset \mathcal{N}} \mathcal{X}(\mathcal{N}')$ defined in Appendix~\ref{app:almost_global}, which can be made arbitrarily small by reducing $\epsilon$.

Therefore, we can prove the existence of the almost semi-globally rapidly stable limit cycle for the normalized network with a sufficiently small $\epsilon$.
Since, for fixed $\epsilon$, we have finite velocity in the entire region, the shape and the rapid stability of the limit cycle are preserved.

\section{Fast-slow behavior in the rest of the region $\mathcal{D} \setminus (\mathcal{S} \cup \mathcal{UF} \cup \mathcal{F})$}\label{app:almost_global}

To complete the illustration on the network that we obtain by the time scale separation for the $\epsilon = 0$ limit, we will detail the fast-slow behavior in the rest of the region $\mathcal{D} \setminus (\mathcal{S} \cup \mathcal{UF} \cup \mathcal{F})$.
Recall that any trajectory that starts from the slow region $\mathcal{S}$ moves to the fast region $\mathcal{F}$ passing through the ultrafast region $\mathcal{UF}$, and any trajectory that starts from the fast region $\mathcal{F}$ moves to the slow region $\mathcal{S}$ passing through the ultrafast region $\mathcal{UF}$ after possibly a finite number of iteration $\mathcal{F} \to \mathcal{UF} \to \mathcal{F}$.

Now, assume without loss of generality that 
$$\mathcal{D}_i = [\underline{x}_i(0), \overline{x}_i(M_i)] \times [v_i^\text{sm}(0, \overline{x}_i(M_i)), v_i^\text{lg}(M_i, \underline{x}_i(0))], \,\,\, \forall i.$$
Then, the remaining region can be partitioned as 
$$\bigcup_{\mathcal{N}' \subset \mathcal{N}} \mathcal{M}(\mathcal{N}'),$$
where $\mathcal{M}(\mathcal{N}')$ is the set of all points ${\rm col}(x_1, v_1, \dots, x_N, v_N)$ such that
\begin{align*}
0 &= f_i(x_i, v_i) + g_i(\overline{E}_i - v_i)\sum_{j \in \mathcal{N}_i}\alpha_{ij}S_{ij}(v_j) \\
&=: f_i(x_i, v_i) + g_i(\overline{E}_i - v_i)d_i(v_1, \dots, v_N), \quad i \in \mathcal{N},
\end{align*}
and $v_i = v_i^\text{md}(g_id_i(v_1, \dots, v_N), x_i)$ if and only if $i \notin \mathcal{N}'$, i.e., $v_i = v_i^\text{sm}(g_id_i(v_1, \dots, v_N), x_i)$ or $v_i = v_i^\text{lg}(g_id_i(v_1, \dots, v_N), x_i)$ for $i \in \mathcal{N}'$, where $v_i^\text{md}(m, x)$ denotes the middle root of $f_i^m(x, v) = 0$ for $x \in (\underline{x}_i(m), \overline{x}_i(m))$ so that $v_i^\text{sm}(m, x) < v_i^\text{md}(m, x) < v_i^\text{lg}(m, x)$.

A similar argument as earlier shows that neurons $i$ in the lower branch or the upper branch, i.e., $i \in \mathcal{N}'$, does not move towards the middle branch, hence any trajectory that starts from $\mathcal{M}(\mathcal{N}')$ can only move to either $\mathcal{S} \cup \mathcal{F}$ or $\mathcal{M}(\mathcal{N}'')$ with $\mathcal{N}'' \supseteq \mathcal{N}'$, passing through the ultrafast region $\mathcal{UF}$.
In particular, note that any trajectory that starts from the set $\mathcal{M}(\mathcal{N}')$ ends up in either $\mathcal{S}$ or 
\begin{align*}
\mathcal{M}^\mathcal{S}(\mathcal{N}'') &:= \{{\rm col}(x_1, v_1, \dots, x_N, v_N) \in \mathcal{M}(\mathcal{N}''): \\
&\,\,\quad\quad\quad v_i = v_i^\text{sm}(g_id_i(v_1, \dots, v_N), x_i), i \in \mathcal{N}''\}
\end{align*}
with some $\mathcal{N}'' \supseteq \mathcal{N}'$, via the ultrafast-fast behavior.

Therefore, in the remaining part of this section, we will show that $\mathcal{M}^\mathcal{S}(\mathcal{N}')$ is partitioned into a measure zero set $\mathcal{X}(\mathcal{N}')$ and the rest $\mathcal{M}^\mathcal{S}(\mathcal{N}') \setminus \mathcal{X}(\mathcal{N}')$ where a trajectory that starts from it moves to either $\mathcal{S}$ or $\mathcal{M}^\mathcal{S}(\mathcal{N}'')$ with some $\mathcal{N}'' \supsetneq \mathcal{N}'$, via the ultrafast-fast behavior.
In particular, let
\begin{align*}
\mathcal{X}(\mathcal{N}') &:= \{{\rm col}(x_1, v_1, \dots, x_N, v_N) \in \mathcal{M}^\mathcal{S}(\mathcal{N}'):\\
&\,\,\,\quad\quad\quad\quad\quad\quad \exists i \notin \mathcal{N}' \text{ s.t. } x_i = x_{i, \infty}(v_i)\}, \\
\mathcal{S}^\mathcal{M}(\mathcal{N}') &:= \{{\rm col}(x_1, v_1, \dots, x_N, v_N) \in \mathcal{M}^\mathcal{S}(\mathcal{N}') : \\
&\quad\quad\quad\quad\quad\quad\quad\quad\quad\quad v_i \le E_i^\text{th}, \forall i \notin \mathcal{N}'\}, \\
\mathcal{M}^{\mathcal{S},left}(\mathcal{N}') &:= \{{\rm col}(x_1, v_1, \dots, x_N, v_N) \in \mathcal{M}^\mathcal{S}(\mathcal{N}') : \\
&\,\,\,\quad\quad v_i \le E_i^\text{th} \text{ or } x_i < x_{i, \infty}(v_i) , \forall i \notin \mathcal{N}'\}, \\
\mathcal{M}^{\mathcal{S}, right}(\mathcal{N}') &:= \mathcal{M}^\mathcal{S}(\mathcal{N}') \setminus (\mathcal{X}(\mathcal{N}')  \cup \mathcal{M}^{\mathcal{S}, left}(\mathcal{N}')).
\end{align*}
Then, $\mathcal{X}(\mathcal{N}')$ is a measure zero set and any trajectory that starts from $\mathcal{M}^{\mathcal{S}, left}(\mathcal{N}') \setminus \mathcal{S}^\mathcal{M}(\mathcal{N}')$ moves either to $\mathcal{S}^\mathcal{M}(\mathcal{N}')$ via the fast behavior or to $\mathcal{M}^\mathcal{S}(\mathcal{N}'')$ via the ultrafast-fast behavior with some $\mathcal{N}'' \supsetneq\mathcal{N}'$.
On the other hand, any trajectory that starts from $\mathcal{M}^{\mathcal{S}, right}(\mathcal{N}')$ moves to $\mathcal{M}^\mathcal{S}(\mathcal{N}'')$ via the ultrafast-fast behavior with some $\mathcal{N}'' \supsetneq\mathcal{N}'$.
Finally, any trajectory that starts from $\mathcal{S}^\mathcal{M}(\mathcal{N}')$ moves either to $\mathcal{M}^{\mathcal{S}, right}(\mathcal{N}')$ via the slow behavior or to $\mathcal{M}^\mathcal{S}(\mathcal{N}'')$ via the slow-ultrafast behavior with some $\mathcal{N}'' \supsetneq \mathcal{N}$.

\end{document}